    \documentclass[acmsmall,screen]{acmart}
    
    \AtBeginDocument{%
      \providecommand\BibTeX{{%
        \normalfont B\kern-0.5em{\scshape i\kern-0.25em b}\kern-0.8em\TeX}}}
    
    \setcopyright{rightsretained}
    \copyrightyear{2021}
    \acmYear{2022}
    \acmDOI{10.1145}
    \acmVolume{1}
    \acmNumber{1}
    \acmArticle{1}
    \acmMonth{3}

\usepackage{threeparttable}
\usepackage{array,tabularx}
\usepackage{url}
\usepackage{graphicx}
\usepackage{color}
\usepackage[linesnumbered,ruled,vlined]{algorithm2e} 
\usepackage{siunitx}
\usepackage{subcaption}

\makeatletter
\usepackage{tikz}
\usetikzlibrary{calc}
\newcommand*\circled[1]{\tikz[baseline=(char.base)]{
    \node[shape=circle, draw, inner sep=0.5pt, 
        minimum height={\f@size*1.1},] (char) {\vphantom{WAH1g}#1};}}
\makeatother

\usepackage{soul}
\usepackage{xspace}
\usepackage{tikz}
\usetikzlibrary{patterns}
\usepackage{pgfplots}
\usepackage{pgfplotstable}

\newcommand{\comet}{\emph{CoMeT}\xspace}
\newcommand{\revision}{\textcolor{black}}
\newcommand{\SimCtrl}{\emph{SimulationControl}\xspace}
\newcommand{\hv}{\emph{HeatView}\xspace}
\pgfplotsset{compat=1.13,
	/pgfplots/ybar legend/.style={
		/pgfplots/legend image code/.code={%
			\draw[##1,/tikz/.cd,yshift=-0.25em]
			(0cm,0cm) rectangle (3pt,0.6em);},
	},
	scale only axis,  
	xlabel near ticks,
	ylabel near ticks,
	xlabel style={font=\footnotesize,align=center},
	ylabel style={font=\footnotesize,align=center},
	xlabel style={yshift=1.5mm},
	ylabel style={yshift=-1.5mm},
	x tick label style={font=\footnotesize},
	y tick label style={font=\footnotesize},
	legend style={
		font={\footnotesize},
		draw=none,
		fill=none,
		/tikz/every even column/.append style={column sep=2mm},
	},
	legend cell align={left},
	max space between ticks=20pt  
}

\tikzset{
    core_color/.style={
        blue,
    },
    core_bar/.style={
        fill=blue,
    },
    mem_color/.style={
        orange!80!red,
    },
    mem_bar/.style={
        preaction={fill=orange!80!red!80!white},
    },
    other_color/.style={
        black!70!white,
    },
    other_bar/.style={
        preaction={fill=black!70!white},
    },
}

\usepackage{hyperref}
\hypersetup{
    colorlinks,
    citecolor=black,
    filecolor=black,
    linkcolor=black,
    urlcolor=black
}

\SetCommentSty{mycommfont}
\SetKwComment{Comment}{$\triangleright$\ }{}
\usepackage{blindtext}
\usepackage{comment}
\begin{document}
%
\ifdefined\IEEEformat
    \title{CoMeT: An Integrated Interval Thermal Simulation Toolchain for 2D, 2.5D, and 3D Processor-Memory Systems}

\else
    \title[CoMeT: Integrated \underline{Co}re and \underline{Me}mory \underline{T}hermal Simulation Toolchain]{CoMeT: An Integrated Interval Thermal Simulation Toolchain for 2D, 2.5D, and 3D Processor-Memory Systems}

\author{Lokesh Siddhu}
\affiliation{%
  \institution{Department of CSE, Indian Institute of Technology Delhi}
  \city{New Delhi}
  \country{India}}
\email{siddhulokesh@cse.iitd.ac.in}

\author{Rajesh Kedia}
\affiliation{%
  \institution{Department of CSE, Indian Institute of Technology Hyderabad}
  \city{Telangana}
  \country{India}}
  \authornote{This work was started during his affiliation with Khosla School of IT, IIT Delhi, India.}
\email{rkedia@cse.iith.ac.in}

\author{Shailja Pandey}
\affiliation{%
  \institution{Department of CSE, Indian Institute of Technology Delhi}
  \city{New Delhi}
  \country{India}}
\email{shailjapandey@cse.iitd.ac.in}

\author{Martin Rapp}
\affiliation{%
  \institution{Chair for Embedded System (CES), Karlsruhe Institute of Technology (KIT)}
  \city{Karlsruhe}
  \country{Germany}}
\email{martin.rapp@kit.edu}

\author{Anuj Pathania}
\affiliation{%
  \institution{Informatics Departments, University of Amsterdam}
  \city{Amsterdam}
  \country{Netherlands}}
\email{a.pathania@uva.nl}

\author{J\"org Henkel}
\affiliation{%
  \institution{Chair for Embedded System (CES), Karlsruhe Institute of Technology (KIT)}
  \city{Karlsruhe}
  \country{Germany}}
\email{henkel@kit.edu}

\author{Preeti Ranjan Panda}
\affiliation{%
  \institution{Department of CSE, Indian Institute of Technology Delhi}
  \city{New Delhi}
  \country{India}}
\email{panda@cse.iitd.ac.in}

\fi

%

\ifdefined\IEEEformat
    \markboth{Journal of \LaTeX\ Class Files,~Vol.~14, No.~8, August~2015}%
    {Shell \MakeLowercase{\textit{et al.}}: Bare Demo of IEEEtran.cls for IEEE Journals}
%
\fi



\ifdefined\IEEEformat
 \else
    \renewcommand{\shortauthors}{Siddhu et al.}
\fi

\ifdefined\IEEEformat
    \maketitle
\fi



\begin{abstract}
\ifdefined\IEEEformat
\else
    \hrulefill
    
\fi
 
Processing cores and the accompanying main memory working in tandem enable the modern processors. Dissipating heat produced from computation remains a significant problem for processors. Therefore, the thermal management of processors continues to be an active subject of research. Most thermal management research is performed using simulations, given the challenges in measuring temperatures in real processors. Fast yet accurate interval thermal simulation toolchains remain the research tool of choice to study thermal management in processors at the system level. However, the existing toolchains focus on the thermal management of cores in the processors since they exhibit much higher power densities than memory. 

The memory bandwidth limitations associated with 2D processors lead to high-density 2.5D and 3D packaging technology. 2.5D packaging technology places cores and memory on the same package. 3D packaging technology takes it further by stacking layers of memory on the top of cores themselves. These new packagings significantly increase the power density of the processors, making them prone to overheating. Therefore, mitigating thermal issues in high-density processors (packaged with stacked memory) becomes even more pressing.  However, given the lack of thermal modeling for memories in existing interval thermal simulation toolchains, they are unsuitable for studying thermal management for high-density processors.

To address this issue, we present the first integrated \underline{Co}re and \underline{Me}mory interval \underline{T}hermal simulation toolchain called \comet.
\comet comprehensively supports thermal simulation of high- and low-density processors corresponding to four different core-memory (integration) configurations -- off-chip DDR memory, off-chip 3D memory, 2.5D, and 3D. 
\comet supports several novel features that facilitate overlying system research.
\comet adds only an additional $\sim$5\% simulation-time overhead compared to an equivalent state-of-the-art core-only toolchain.
The source code of \comet has been made open for public use under the {\em MIT} license.
\end{abstract}

\ifdefined\IEEEformat
\else
\fi

\ifdefined\IEEEformat
\else


\fi

\ifdefined\IEEEformat
    \IEEEpeerreviewmaketitle
\else
    \maketitle
    \hrulefill
\fi
%

\section{Introduction}
\label{sec:introduction}

Processing cores and the accompanying main memory working together make the modern processor work. 
It is common to fabricate the cores and memory separately on different packages using 2D packaging technology and then connect them using off-chip interconnects. 
However, the limited bandwidth of the interconnect often becomes the performance bottleneck in the 2D processor. 
Recent advances in semiconductor manufacturing have enabled high-density integration of core and memory wherein the designers place them on the same package using 2.5D packaging technology to improve bandwidth. 
Designers can now  stack memory and core on top of each other as layers using 3D packing technology for several magnitudes increase in bandwidth~\cite{loh2007processor}.
These advances enable the next generation of high-performing high-density 2.5D and 3D processors.

The tighter (and vertical) integration of core and memory into a single package results in the power of both core and memory getting channeled in the same package. 
However, there is not much increase in the package’s corresponding surface area. 
Consequently, the integration significantly increases the power density of the processor. 
Therefore, these high-density processors (packaged with stacked memory) face even more severe thermal issues than low-density 2D processors~\cite{coudrain2016experimental}. 
Promising as they are, the thermal issues associated with high-density processors prevent them from going mainstream.
Therefore, thermal management for high-density processors is now an active research subject~\cite{Hajkazemi2017, Lo2016, siddhu2019predictncool,siddhu2020leakage}.

However, the availability of thermal sensors in real-processors is limited, and they often lack the temporal and spatial resolutions needed for thermal management research.
Given the challenges involved in measuring temperatures in real-world processors, thermal simulations play an essential role in enabling  thermal management research.
However, due to the lack of better open-source tools, existing works on thermal management of high-density processors and most works on thermal management of low-density processors  are based on  in-house trace-based simulators~\cite{cao2019survey}. 
Recent advances in Electronic Design Automation~(EDA) have enabled detailed core-only interval thermal simulations using sophisticated open-source toolchains~\cite{pathania2018hotsniper,rohith2018lifesim, Hankin:2021:Hotgauge}.
Trace-based simulation relies on first collecting traces (performance, power) of each application running in isolation. 
It then performs segregated temperature simulations on the merged (independent) traces.
In contrast, an interval-based simulation executes all applications in parallel, allowing it to consider contention on shared resources.
\revision{The following motivational example shows that interval simulations are more detailed and accurate than trace-based simulations.}

\begin{figure}
    \centering
    \pgfplotsset{
        motivational_temperature/.style={
            width=6.0cm,
            height=4.2cm,
            xmin=0,
            xmax=185,
            xlabel={Time (ms)},
            ylabel={Temperature ($^\circ$C)},
            ymajorgrids,
            legend columns=2,
            legend style={
                at={(0.5,1)},
                anchor=south,
            },
        },
        motivational_frequency/.style={
            width=5cm,
            height=1.7cm,
            xmin=0,
            xmax=440,
            ymin=0,
            xlabel={Time (ms)},
            ylabel={Performance (IPS)},
            ymajorgrids,
            legend columns=2,
            legend style={
                at={(1,1)},
                anchor=south east,
            },
        },
        trace/.style={
            red
        },
        interval/.style={
            blue
        },
        swaptions/.style={
            orange
        },
        ocean/.style={
            green!50!black
        }
    }
    \begin{tikzpicture}
        \begin{axis}[motivational_temperature,name=temp_axis,at={(0,0)},anchor=east]
            \addplot[interval]
                table[x=t,y=interval,col sep=comma]
                {data/motivational_temperature.csv};
            \addplot[trace]
                table[x=t,y=trace,col sep=comma]
                {data/motivational_temperature.csv};
            \legend{Interval Simulation, Trace-based Simulation}
            
            \coordinate (top) at (rel axis cs:0,1);
            \coordinate (bottom) at (rel axis cs:0,0);
            \coordinate (traceend) at (axis cs:152,0);
            \coordinate (intervalend) at (axis cs:181,0);

            \draw [densely dashed] (traceend |- top) to (traceend |- bottom);
            \draw [densely dashed] (intervalend |- top) to (intervalend |- bottom);

            \coordinate (perflabely) at (axis cs:0,75.5);
            \draw [-latex]
                (intervalend |- perflabely) to
                node [below=0.5mm, font=\footnotesize,circle,draw,fill=white,inner sep=1pt] {I}
                (traceend |- perflabely);

            \coordinate (tracepeak) at (axis cs: 142,78.18);
            \coordinate (intervalpeak) at (axis cs: 172,77.01);

            \draw [densely dashed]
                ([xshift=-2mm]tracepeak) to
                ([xshift=+2mm]intervalpeak |- tracepeak);
            \draw [densely dashed]
                ([xshift=-2mm]tracepeak |- intervalpeak) to
                ([xshift=+2mm]intervalpeak);
            
            \coordinate (templabelx) at (axis cs: 160,0);
            \draw [-latex]
                (templabelx |- intervalpeak) to
                node [right=0.5mm, font=\footnotesize,circle,draw,fill=white,inner sep=1pt] {II}
                (templabelx |- tracepeak);
        \end{axis}

        \begin{axis}[
                motivational_frequency,
                at={(15mm,4mm)},
                anchor=south west,]
            \addplot[swaptions]
                table[x=t,y expr=\thisrow{ips_swaptions},col sep=comma]
                {data/motivational2_4GHz.csv};
            \addplot[ocean]
                table[x=t,y expr=\thisrow{ips_ocean},col sep=comma]
                {data/motivational2_4GHz.csv};
            \legend{\emph{swaptions},\emph{ocean.ncont}}
            
            \coordinate (top) at (rel axis cs:0,1);
            \coordinate (bottom) at (rel axis cs:0,0);
            \coordinate (swaptions_poi_high_freq) at (axis cs:100,6e9);
            \coordinate (ocean_poi_high_freq) at (axis cs:100,2e9);
            \draw [densely dashed] (swaptions_poi_high_freq |- top) to (ocean_poi_high_freq |- bottom);
            
            \node [font=\footnotesize,inner sep=1pt,anchor=north,fill=white]
                at (rel axis cs:0.6,0.95)
                {$f=4.0\,$GHz};
        \end{axis}

        \begin{axis}[
                motivational_frequency,
                name=perf_axis,
                at={(15mm,-4mm)},
                anchor=north west,]
            \addplot[swaptions]
                table[x=t,y expr=\thisrow{ips_swaptions},col sep=comma]
                {data/motivational2_1GHz.csv};
            \addplot[ocean]
                table[x=t,y expr=\thisrow{ips_ocean},col sep=comma]
                {data/motivational2_1GHz.csv};

            \coordinate (top) at (rel axis cs:0,1);
            \coordinate (bottom) at (rel axis cs:0,0);
            \coordinate (swaptions_poi_low_freq) at (axis cs:372,1.6e9);
            \coordinate (ocean_poi_low_freq) at (axis cs:227,0.9e9);
            \coordinate (mid) at (rel axis cs:0,0.5);
            \draw [densely dashed] (swaptions_poi_low_freq |- top) to (swaptions_poi_low_freq |- mid) to (ocean_poi_low_freq |- mid) to (ocean_poi_low_freq |- bottom);
            
            \node [font=\footnotesize,inner sep=1pt,anchor=north,fill=white]
                at (rel axis cs:0.6,0.959)
                {$f=1.0\,$GHz};
        \end{axis}

        \draw [latex-latex,shorten >=1mm,shorten <=1mm]
            (swaptions_poi_high_freq) to[out=-60,in=120]
            node [font=\footnotesize,above right,pos=0.3,align=left,inner sep=1pt,fill=white] {Same Number\\of Executed Instructions}
            (swaptions_poi_low_freq);
        \draw [latex-latex,shorten >=1mm,shorten <=2mm]
            (ocean_poi_high_freq) to[out=-70,in=110] (ocean_poi_low_freq);

        \node[font=\footnotesize,anchor=north,align=center]
            at ([yshift=-7mm]temp_axis.south)
            {(a) Shared resource contention\\between two instances of \emph{ocean.ncont}.};
        \node[font=\footnotesize,anchor=north,align=center]
            at ([yshift=-7mm]perf_axis.south)
            {(b) Sensitivity of the performance\\of \emph{swaptions} and \emph{ocean.ncont} on the frequency.};
    \end{tikzpicture}
    \caption{\revision{
        Trace-based simulation cannot accurately simulate multiple parallel applications.
        (a)~Execution traces obtained in isolation cannot model shared resource contention, resulting in an overestimation of the performance (I) and temperature (II).
        (b)~Parallel traces cannot overcome this limitation in the presence of DVFS, as the performance of different applications depends differently on the frequency. This would require traces of all combinations of applications, at all V/f levels, and at all relative shifts of applications.
    }}
    \label{fig:motivational}
\end{figure}
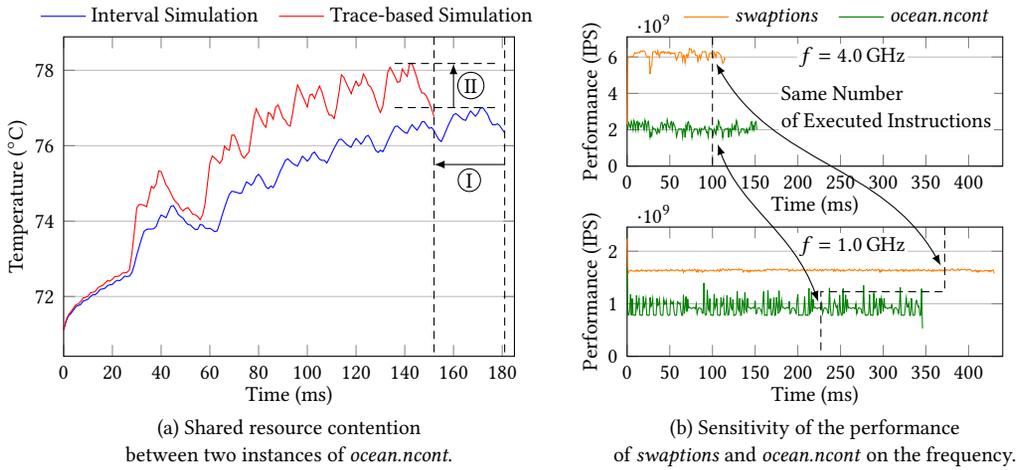

\revision{
\textbf{Motivational Example:} Figure~\ref{fig:motivational}(a) shows a scenario in which two instances of \emph{SPLASH-2 ocean.ncont} are executing in parallel.
Both instances of \emph{ocean.ncont} compete for DRAM bandwidth, which leads to a performance reduction of 16\% due to stall cycles; trace-based simulation (I) cannot capture this effect.
In addition, stall cycles reduce the power consumption and consequently the temperature.
Therefore, trace-based simulation overestimates the temperature (II).
One can overcome this overestimation by obtaining traces for all combinations of applications, but such an approach might already be prohibitive.
Dynamic Voltage Frequency Scaling (DVFS) technology in processors further aggravates the problem.
Scaling V/f levels affects the performance of memory- and compute-intensive applications differently.
Figure~\ref{fig:motivational}(b) shows the traces of the compute-intensive \emph{PARSEC swaptions} and the memory-intensive \emph{SPLASH-2 ocean.ncont} at 4\,GHz~(top) and 1\,GHz~(bottom).
The points in their execution the two applications reach (after 100\,ms) at 4\,GHz is different from the points they reach when operating at 1\,GHz. 
Therefore, the trace one obtains at a constant of 1\,GHz can not be used to continue a simulation that switches from 4\,GHz to 1\,GHz after 100\,ms. 
The trace-based simulation would require traces of all combinations of applications at all V/f levels and all relative shifts of applications. The collection of all these traces is practically infeasible.}

Cycle-accurate simulations~\cite{binkert2011gem5} are more accurate than interval simulations.
However, they are also extremely slow and difficult to parallelize (often single-threaded).
Cycle-accurate simulations are quintessential to test the accuracy of new micro-architecture designs, which designers can do in a limited number of processor cycles.
In system research, on the other hand, we are required to simulate multiple (many) processors simultaneously for time measured in minutes (hours) rather than cycles to reproduce the necessary system-level behavior.
For example, a single-core processor running at 1 GHz goes through a billion processor cycles every second.
Cycle-accurate simulations are therefore too slow for system-level research.
Interval simulations are several magnitudes faster than cycle-accurate simulations and provide a good trade-off between simulation speed and accuracy. 
Interval simulations are therefore best suited for system-level research that requires simulation of a multi-/many-core processor for a long duration but with high fidelity. 
However, existing interval thermal simulation toolchains do not model the main memory and cannot be used to study the high-density processors wherein core and memory are tightly integrated and thermally coupled.

\begin{figure}[t]
\centering
      \includegraphics[width=0.72\linewidth]{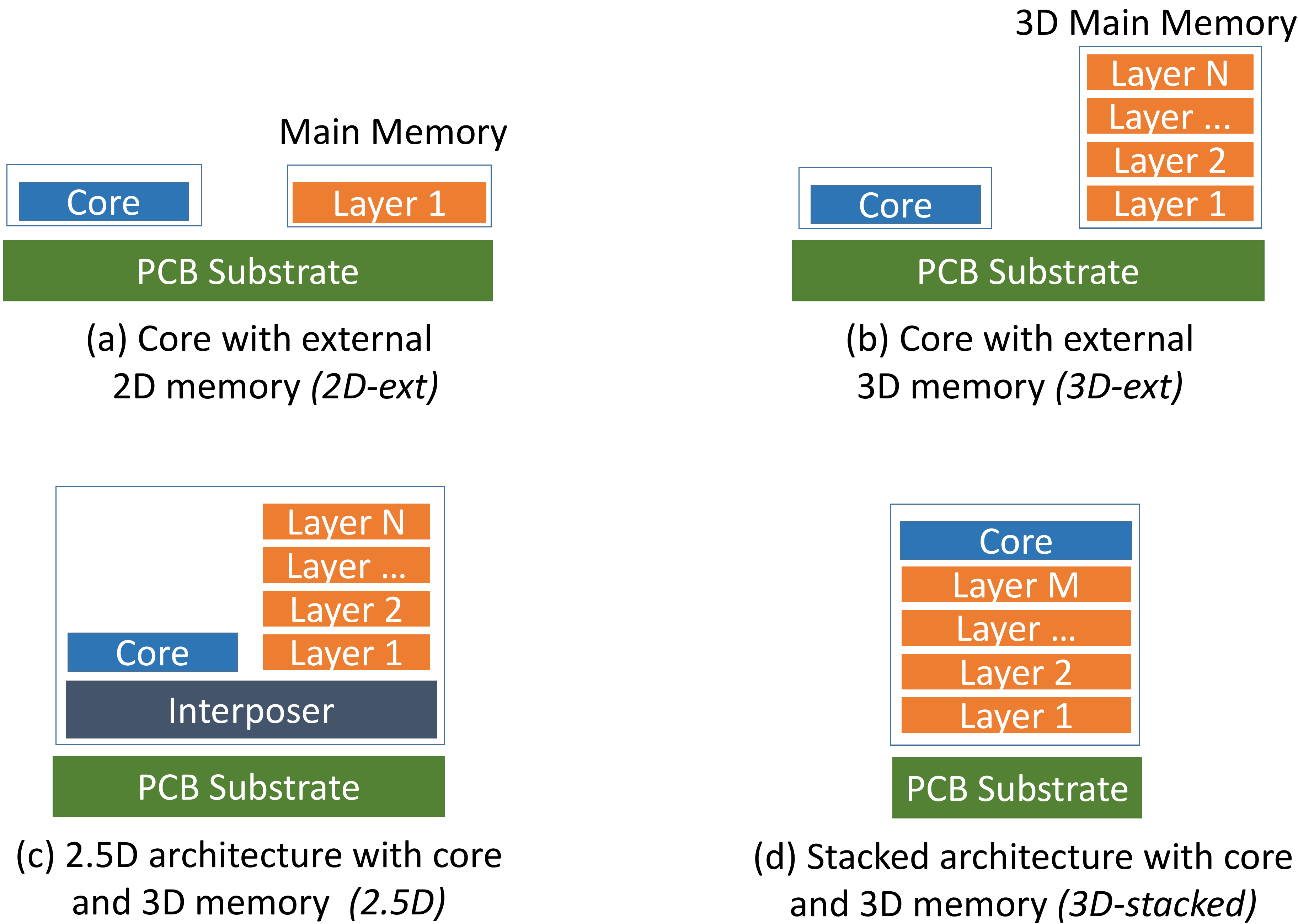}
      \caption{Various core-memory configurations. Core also includes caches \revision{and can have multiple layers as well.} 
      }
      \label{fig:SysMemArch}
\end{figure}

In this work, we present the first interval thermal simulation toolchain, called \comet, that holistically integrates both core and memory.
\comet provides performance, power, and temperature values at regular user-defined intervals (epochs) for core and memory. 
The support for thermal interval simulation for both core and memory using the \comet toolchain comes at only $\sim$5\% additional simulation-time overhead over {\em HotSniper}~\cite{pathania2018hotsniper} (state-of-the-art thermal interval simulation toolchain for core-only simulations). 
\comet enables users to evaluate and analyze run-time thermal management policies for various core-memory (integration) configurations as shown in Figure~\ref{fig:SysMemArch}~\cite{coudrain2016experimental,stow2016cost, sodani2015knights, hassan2015near, park2021high, loh20083d}.
Figure~\ref{fig:SysMemArch}(a) shows a conventional but the most common configuration with cores and 2D DRAM on separate packages. 
Figure~\ref{fig:SysMemArch}(b) replaces the 2D DRAM with a 3D memory for faster data access. Figure~\ref{fig:SysMemArch}(c) further bridges the gap between core and 3D memory by putting them side by side within a package. Figure~\ref{fig:SysMemArch}(d) advances the integration by stacking cores over the 3D memory to reduce data access delays further. 
We refer to configurations shown in Figures~\ref{fig:SysMemArch}(a),~\ref{fig:SysMemArch}(b),~\ref{fig:SysMemArch}(c), and~\ref{fig:SysMemArch}(d) as {\em 2D-ext}, {\em 3D-ext}, {\em 2.5D}, and {\em 3D-stacked}, respectively.

We see \comet primarily as a tool for system-level thermal management research. \comet, therefore, comes equipped with several features that facilitate system research.
\comet ships with \textit{SchedAPI} Application Programming Interface (API) library, which the users can use to implement their custom thermal (resource) management policies.
We also develop and integrate \hv into \comet, which generates a representative video of the thermal simulation for a quick human-comprehensible visual analysis.
It also contains an integrated floorplan generator.
\comet has an extendable automatic build verification~(smoke) test suite that checks critical functionalities across various core-memory configurations and their underlying architectural parameters for quick validation of code edits. 
We develop and integrate the \SimCtrl framework in \comet, using which the users can run simulations for various workloads and configurations in batch mode. 


Using \comet, we also illustrate the thermal patterns for different core-memory configurations using benchmarks from several diverse benchmark suites. These experiments helped us develop many insights into the thermal interactions of cores and memory and their influence on each other's temperatures. 
We also present a thermal-aware scheduling case study wherein we simulate operations of the default {\em on-demand} Governor~\cite{pallipadi2006ondemand} from {\em Linux} operating in conjunction with a 
Dynamic Thermal Management~(DTM) on a 3D stacked processor.
We make several new interesting thermal observations through our case study.
In the same spirit, we envision other researchers will also identify several new thermal behaviors for existing and upcoming core-memory configurations using \comet. Consequently, \comet will enable them to propose and evaluate novel thermal management policies for these configurations.

In particular, we make the following key contributions in this work. 
\begin{enumerate}

    \item We introduce an open-source interval thermal simulation toolchain, called \comet, that holistically integrates core and memory. 
     It supports the simulation of multi-/many-core processors in several different core-memory configurations. 
    
    \item We describe several novel features in \comet that facilitate system-level thermal management research in processors.
    
    \item We perform thermal analysis of different core-memory configurations using \comet and present new interesting thermal observations. 
    We also highlight the suitability of \comet for studying thermal-aware system scheduling via a case study.

\end{enumerate}

\textbf{Open Source Contribution:} The source code for \comet is released under {\em MIT} license for unrestricted use and is available for download at \href{https://github.com/marg-tools/CoMeT}{https://github.com/marg-tools/CoMeT}. 


\section{Background and Related Work}
\label{sec:related_work}

Thermal-aware design of computing systems has been a significant area of research since the early 2000s. With the current technology nodes, the phenomenon of dark silicon (not being able to use the entire chip simultaneously due to thermal issues)~\cite{pathania2018hotsniper, henkel2015darksilicon} is becoming prominent. It is driving the need for fine-grained thermal management to respect the thermal limits of the system. Multi-/many-core processors exhibit unbalanced temperatures and distributed thermal hotspots, making thermal management non-trivial~\cite{khdr2014mdtm}. Various works have addressed thermal management for cores using different techniques~\cite{Huang2000, lasbouygues2007temperature, calimera2008thermal, bao2009line, homayoun2010relocate, kumar2010neural, bailis2011dimetrodon, ayoub2013cometc, liu2013layout, cox2013thermal, sironi2013thermos, zapater2013leakage, cochran2013thermal, sridhar20133d, juan2014statistical, khdr2014mdtm, zhang2015hotspot, prakash2016improving, bogdan2016power, Zheng2016, wang2017fast, kumar2017fighting, liu2018thermal, pathania2018hotsniper, hmctherm, sadrosadati2019itap, smartboost}.
These works primarily include voltage and frequency scaling, hardware reconfiguration, power and clock gating, and cache throttling as knobs for thermal management.
They propose proactive thermal management policies such as task allocation based on future temperature and reactive thermal management policies such as task migration and scheduling for cores. Also, some works have addressed optimizing multiple metrics such as energy, power, and temperature. To name a few, Huang et al.~\cite{Huang2000} proposed a framework for dynamic management of energy and temperature of core in a unified manner. Khdr et al.~\cite{khdr2014mdtm} proposed a technique that employs centralized and distributed predictors to prevent violating temperature thresholds while maintaining the balance between the temperature of different cores. Zapater et al.~\cite{zapater2013leakage} proposed a temperature- and leakage-aware control policy to reduce the energy consumption of data centers by controlling the fan speed. 


Designing appropriate thermal management policies requires a fast and accurate thermal simulator for quick evaluation,  which has resulted in the development of various open-source thermal simulators such as {\em 3D-ICE}~\cite{sridhar20133d} and {\em HotSpot}~\cite{zhang2015hotspot}. Such thermal simulators~\cite{zhang2015hotspot, sridhar20133d, dramsim3, sultan2020fast, sultan2021variability, wang2017fast} use floorplan and power traces as inputs and generate temperature values as output.
{\em 3D-ICE}~\cite{sridhar20133d} is a thermal simulator having a transient thermal model with microchannel cooling for liquid-cooled ICs. {\em HotSpot}~\cite{zhang2015hotspot} provides a fast and accurate thermal model for transient and steady-state simulations.
Thermal simulators pave the way for an early-stage understanding of potential thermal issues in chips and facilitate studies to understand the implications of different designs and floorplans and develop cooling solutions.
A performance simulator for processors, such as {\em Sniper}~\cite{sniper} or {\em gem5}~\cite{binkert2011gem5}, integrated with a power model (such as {\em McPAT}~\cite{mcpat}) generates the power traces used inside these thermal simulators. 
{\em McPAT}~\cite{mcpat} framework can model the power, area, and timing of processor components. It supports technology nodes ranging from 90\,nm to 22\,nm.  {\em Sniper}~\cite{sniper} is a multi-/many-core performance simulator that uses interval simulation to simulate a system at a higher level of abstraction than a detailed cycle-accurate simulator. {\em Sniper} achieves several magnitudes faster simulation speeds over a cycle-accurate simulator such as {\em gem5}. {\em Sniper} integrates {\em McPAT} and enables regular monitoring of the processor's power consumption.

\revision{{Looking at the memory part, \em DRAMsim3}~\cite{dramsim3} and {\em HMCTherm}~\cite{hmctherm} are cycle-accurate DRAM simulators and support thermal simulation.
While {\em DRAMsim3} models 2D and 3D DRAMs, {\em HMCTherm} models 3D memory based on Hybrid Memory Cube (HMC) specification. The detailed cycle-accurate modeling significantly reduces their simulation speed and makes them unsuitable for integration with existing core-only interval-based performance simulators~\cite{sniper}. Further, {\em DRAMsim3} and {\em HMCTherm} focus on off-chip memories and do not consider novel technologies such as 2.5D or 3D integration of cores and memories. {\em CACTI-3DD}~\cite{cacti} is an architecture-level integrated power, area, and timing modeling framework for new memory technologies such as 3D-stacked memories in addition to commodity 2D DRAM and caches. It enables easier integration with an architectural-level core performance simulator.}

Several works have used trace-based evaluation for core thermal management policies~\cite{cox2013thermal, Zheng2016, liu2018thermal, deshwal2019moos}. Cox et al.~\cite{cox2013thermal} use trace-based simulation using {\em HotSpot} to obtain temperature and perform a thermal-aware mapping of streaming applications on 3D processors. Liu et al.~\cite{liu2018thermal} use a trace-based thermal simulation methodology using power traces generated from {\em gem5} + {\em McPAT} for dynamic task mapping on systems with reconfigurable network-on-chip. 
A thermal-aware design space exploration work, {\em MOOS}~\cite{deshwal2019moos}, generates power traces from {\em gem5} and {\em McPAT} and uses {\em HotSpot} and a custom analytical model for temperature estimation of 3D integrated cores and caches. Such a trace-based approach was sufficient for their work as they did not consider any dynamic policy for thermal management. 
A key limitation of evaluating thermal management policies using trace-based simulations is that they do not feed the temperature impact into the performance simulator. 
This limitation limits the accuracy and scope of the analysis.
Many aspects of thermal management, such as reducing the frequency of heated cores based on temperature or adapting cache partitioning based on temperature, cannot be captured by traces collected in isolation and hence can lead to errors or inaccuracies in the overall evaluation. Further, as motivated in Section~\ref{sec:introduction}, an infeasible number of traces might need to be generated to capture the parameter tuning in both performance and thermal simulators. 

Addressing these issues associated with trace-based simulators requires integrating performance and thermal simulators in a coherent manner. {\em HotSniper} was the first to provide an integrated infrastructure for interval-based performance and thermal simulation of 2D processor cores. {\em HotSniper}~\cite{pathania2018hotsniper} integrates the {\em Sniper} performance simulator with {\em HotSpot} thermal simulator and provides an infrastructure for core-only interval thermal simulations of multi-/many-core processors. {\em HotSniper} enables a feedback path for temperature from the thermal simulator ({\em HotSpot}) to the performance simulator ({\em Sniper}) to help make thermal-aware decisions for thermal management. {\em LifeSim}~\cite{rohith2018lifesim} is another notable example of a similar attempt with an additional focus on thermals-based reliability. Recently released {\em HotGauge}~\cite{Hankin:2021:Hotgauge} integrates {\em Sniper} with {\em 3D-ICE}.

Conventionally, memories have lower power dissipation and thus induce lower heating (compared to high-frequency cores~\cite{bircher2008analysis}), thereby requiring limited thermal management. Therefore, prior works such as {\em HotSniper} supported thermal analysis only for cores. With increasing memory bandwidth requirements of applications, high-density {\em 3D-ext}, {\em 2.5D}, and {\em 3D-stacked} processors are becoming popular but face severe thermal issues~\cite{Hajkazemi2017, Lo2016}. Furthermore, high-density processors (and memories within) have significant leakage power dissipation that increases with temperature and forms a positive feedback loop between leakage power and temperature. 
Therefore, in recent times, memory heating in high-density processors has also received significant research attention~\cite{siddhu2020leakage, siddhu2019predictncool}. {\em FastCool}~\cite{siddhu2020leakage} discusses DTM strategies for 3D memory considering the leakage power dissipation in memory and positive feedback loop between leakage power and temperature. {\em PredictNCool}~\cite{siddhu2019predictncool} proposes a proactive DTM policy for 3D memories using a lightweight steady-state temperature predictor. \revision{
Instead of using detailed command level 3D memory models~\cite{dramsim3, hmctherm}, both {\em FastCool} and {\em PredictNCool} obtain energy-per-access from {\em CACTI-3DD}~\cite{cacti} to derive memory power based on access traces and used {\em HotSpot} for thermal simulation in a trace-based methodology.
While they use the same tools (CACTI-3DD, Sniper, and HotSpot) used in \comet, their evaluation suffers from the already discussed limitations of a trace-based simulation. Moreover, such a setup cannot provide dynamic feedback to cores, limiting its scope and accuracy.}

\section{COMET: Integrated Thermal Simulation for Cores and Memories} 
\label{sec:proposal}

\comet integrates a performance simulator~({\em Sniper}~\cite{sniper}), a power model for core~({\em McPAT}~\cite{mcpat}), a power model for memory~({\em CACTI}~\cite{cacti}), and a thermal simulator ({\em HotSpot}~\cite{zhang2015hotspot}) to perform an integrated interval performance, power, and thermal simulation for cores and memories. It also provides many other useful features and utilities to handle multiple core-memory configurations, thermal management, floorplan generation, etc., within the framework. 
We present the proposed \comet tool flow and features in this section.


\subsection{\comet Tool Flow}
\label{sec:ToolFlow}

We first provide an overview of the \comet toolchain and then explain each block in detail.

\subsubsection{Overview}
\label{sec:OverviewToolFlow}

\begin{figure*}[t]
\centering
    \begin{minipage}[t]{\linewidth}
      \centering
      \includegraphics[width=\linewidth]{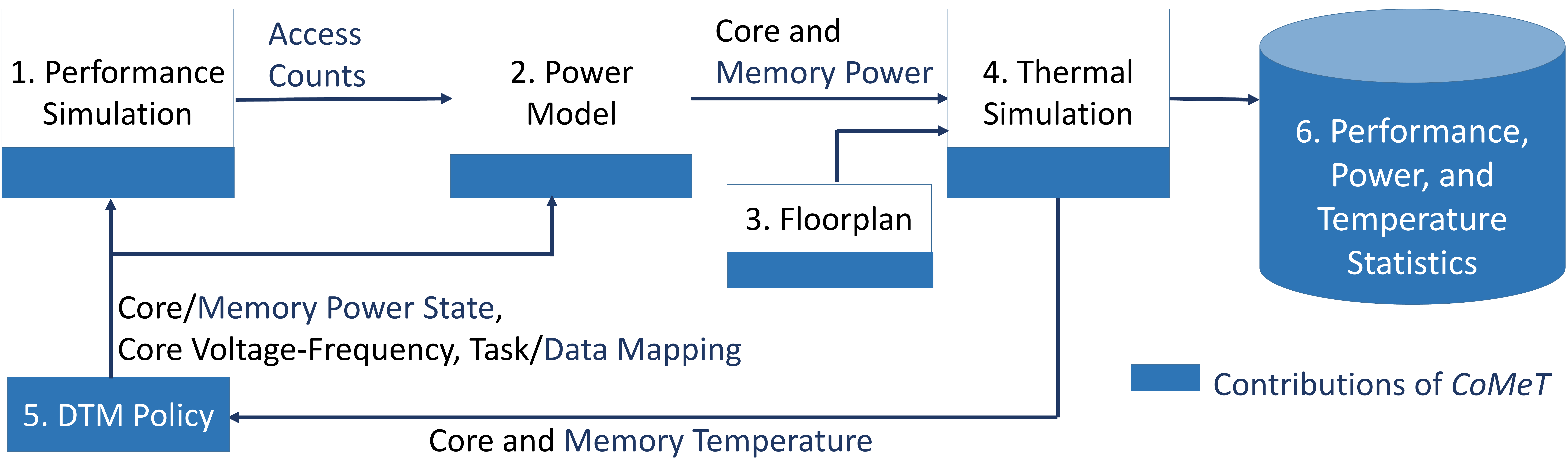}
      \caption{Overview of \comet Flow}
      \label{fig:Tool-Flow}
    \end{minipage}
\end{figure*}

Figure~\ref{fig:Tool-Flow} shows an overall picture of the \comet toolchain. Components in blue indicate the key contributions of \comet.
The toolchain first invokes an \circled{1} interval performance simulator~(e.g., {\em Sniper}~\cite{sniper}) to simulate a workload and tracks access counts to various internal components such as execution units, caches, register files, etc. We also extended the existing performance simulator to monitor memory access counts.
The access counts are accumulated and passed to the \circled{2} power model at every epoch (e.g., \SI{1}{\milli\second}). The power model (e.g., {\em McPAT}~\cite{mcpat}) calculates the core and memory power during the corresponding epoch, which toolchain then feeds along with the chip \circled{3} floorplan to a \circled{4} thermal simulator (e.g., {\em HotSpot}~\cite{zhang2015hotspot}) for calculating core and memory temperature. 
Depending upon the type of core-memory configuration, thermal simulation of core and memory can occur separately (Figures~\ref{fig:SysMemArch}(a), (b)) using two different invocations of the thermal simulator or together (Figures~\ref{fig:SysMemArch}(c), (d)) using a single invocation.
As shown in Figure~\ref{fig:Tool-Flow}, the toolchain provides the core and memory temperatures as inputs to the \circled{5} DTM policy. If the temperature exceeds a threshold, the DTM policy will invoke knobs (e.g., changing the core and memory power state, operating voltage/frequency, task/data mapping, etc.) to manage the temperature. Such knobs would affect the performance simulation, and the above process repeats until the end of the workload.
Once the simulation is complete, \comet collects various metrics such as IPC, cache hit rate, DRAM bandwidth utilization, etc., from the performance simulator, power traces from the power model, and temperature traces from the temperature simulator. These metrics and traces are processed to generate different plots and statistics, enabling easier and more detailed analysis. We provide details of each of these blocks in the following subsection.

\subsubsection{Toolchain Details}
\label{sec:DetailsToolFlow}

\begin{figure*}[t]
\centering
    \begin{minipage}[t]{\linewidth}
      \centering
      \includegraphics[width=\linewidth]{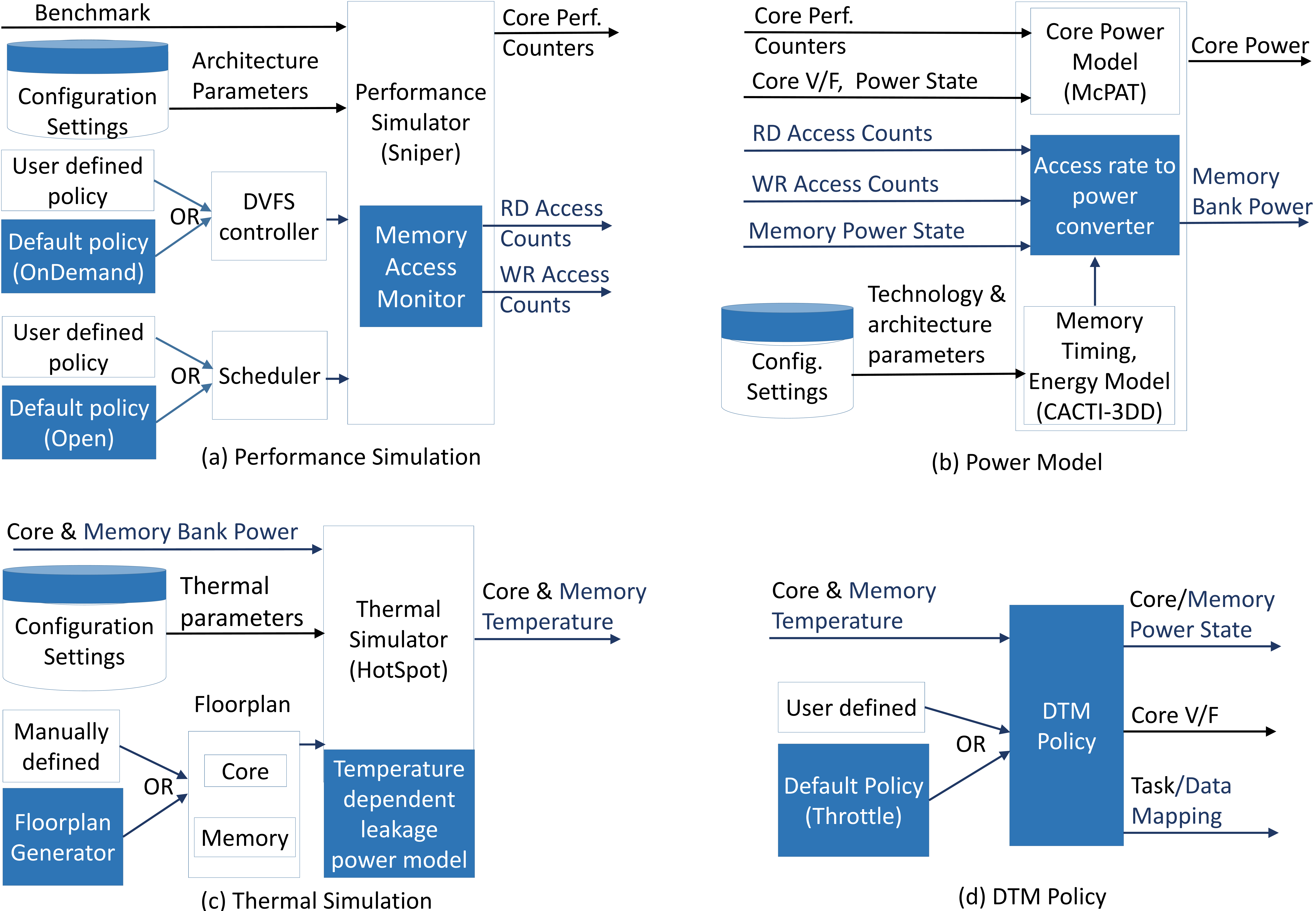}
     \caption{\comet Detailed Flow}
      \label{fig:Detailed-Tool-Flow}
    \end{minipage}
\end{figure*}

Figure~\ref{fig:Detailed-Tool-Flow} shows different blocks of the \comet flow in more detail. Figure~\ref{fig:Detailed-Tool-Flow}(a) illustrates the performance simulation block, which simulates a workload and provides access counts of different core blocks and memory. We updated the {\em Sniper}~\cite{sniper} performance simulator to monitor and accumulate memory read (RD) access and write (WR) access counts (separately for each bank) in each epoch. 
Modern-day cores use various voltage-frequency levels and scheduling strategies to improve performance, increase energy efficiency, and reduce temperature. 
We provide an \emph{ondemand} governor and a scheduler for open systems~\cite{open_system} as default policies that users can suitably modify (more details in Section~\ref{sec:Scheduler}) to jumpstart development. The DVFS controller and the scheduler control various aspects of performance simulation and form inputs to the performance simulator. The user defines different processor architecture parameters such as the number of cores, frequency range, cache sizes, etc., as a part of settings. \comet then provides the settings as inputs to the performance simulation block.

We move now to explain the power model block shown in Figure~\ref{fig:Detailed-Tool-Flow}. \comet uses the access counts generated from the performance simulation block, the power state, and operating voltage/frequency of each core to calculate core and memory power at every epoch. 
The core and memory power is calculated separately for each core and bank, respectively. 
The settings provide various technology parameters (e.g., technology node in nano-meters) and architecture parameters (such as cache attributes, number of functional units, etc.) as inputs to the power model.
\comet calculates the core power using {\em McPAT}~\cite{mcpat}.
\comet first extracts the energy per access for RD and WR operation from a memory modeling tool~({\em CACTI-3DD}~\cite{cacti}) to calculate the memory power.
This energy per access data is used within the access rate to the power converter block to convert the RD and WR access counts of each bank to corresponding dynamic power.

The next block is the thermal simulation block (Figure~\ref{fig:Detailed-Tool-Flow}(c)), which calculates the temperature of individual core and memory banks using their power consumption and the chip floorplan/layer information.
While a user can provide a floorplan file developed manually, we also implemented an automatic floorplan generator to generate the floorplan for various regular layouts (details in Section~\ref{sec:Floorplan})). 
The users provide the floorplan as an input to the thermal simulation block. 
We use a fast and popular thermal simulator called {\em HotSpot}~\cite{zhang2015hotspot}\footnote{\revision{
\comet uses {\em HotSpot} as the thermal simulator, but one can extend it to support any other (more accurate) thermal simulator.  This extension is possible because most thermal simulators (e.g., HotSpot, 3D-ICE) follow similar input-output interfaces but different formats. During each interval, these simulators require a power trace as an input (generated by a performance simulator like Sniper) and generate temperature trace as an output. Therefore, an addition of a trace format converter within \comet should suffice to support different thermal simulators. These simulators also require configuration parameters and floorplan information as inputs which typically remain unchanged during the entire simulation. Thus, different thermal simulators can be supported by generating this information in an appropriate format, either manually or through automation (e.g., \textit{floorplanlib}). A plugin-type integration of various simulators would be useful, and we leave it as future work for now.}}, within \comet, extended to combine the dynamic power with the temperature-dependent leakage power at each epoch. Section~\ref{sec:leakage-aware} presents details of the temperature-dependent leakage-aware modeling. 
The user provides the thermal and tool parameters (epoch time, initial temperatures, config file, etc.) to the thermal simulation block.



As shown in Figure~\ref{fig:Detailed-Tool-Flow}(d), the DTM policy manages temperature by employing a range of actions, such as using low power states, decreasing core voltage-frequency (V/F), changing the task/data mapping, and reducing power density. We provide a default throttle-based scheme (Section~\ref{sec:Scheduler}), which can be used as a template to help users develop and evaluate different thermal management schemes. The DTM policy uses the temperature data provided by the thermal simulation block and controls the power states, V/F settings, or the task/data mapping. 
The performance simulation and power model block make use of these knobs.

After the workload simulation completes, using \SimCtrl, \comet outputs the performance, power, and temperature for various timesteps for both core and memory (not shown in Figure~\ref{fig:Detailed-Tool-Flow}). Such traces are also available in graphical format, enabling a quicker and better analysis. In addition, \hv generates a temperature video showing the thermal map of various cores and memory banks at different time instances. \SimCtrl allows users to run simulations in batch mode (Section~\ref{sec:SimulationControl}) on the input side. We elaborate on the various key features of \comet in the following subsections.




\subsection{Support for Multiple Core-Memory Configurations}
\label{sec:ArchSupport}

In this section, we discuss various core-memory configurations supported in \comet. Today's technology supports integrating the core and memory in a processor (computer system) in multiple ways~\cite{hassan2015near}.
As shown in Figure~\ref{fig:SysMemArch}, we support four different kinds of core-memory configurations in \comet. Designers can package the core and memory separately (Figure~\ref{fig:SysMemArch}(a), (b)) or on the same package (Figure~\ref{fig:SysMemArch}(c), (d)). They also solder the packaged chips onto a Printed Circuit Board~(PCB) for mechanical stability and electrical connectivity.

Off-chip 2D core-memory configurations~\cite{jacob2009memory}, referred to as \textit{2D-ext} in this work, are the most widely used configurations today. In such core-memory configurations, usually, the core has a heat sink for cooling while the DRAM memory is air-cooled (Figure~\ref{fig:SysMemArch}(a)). The \comet toolchain studies thermal issues in such core-memory configurations. In many processors, off-chip 3D memories are becoming popular with the rising need for higher memory bandwidth. However, the increased power density causes thermal issues, requiring a heat sink for memory cooling (Figure~\ref{fig:SysMemArch}(b)). We refer to such core-memory configurations as \textit{3D-ext} in this work. 
\revision{The 3D memory contains a logic core layer (not shown in the figure) at the bottom which manages the routing of requests and data between various layers of the 3D memory.}

The above off-package core-memory configurations (\textit{2D-ext} or \textit{3D-ext}) have a higher interconnect delay. In an alternative core-memory configuration referred to as \textit{2.5D}~\cite{coudrain2016experimental, hassan2015near} (Figure~\ref{fig:SysMemArch}(c)), a 3D memory and 2D core are placed within the same package, thereby reducing the interconnect delay. An interposer~\cite{coudrain2016experimental} acts as a substrate and helps route connections between the memory and core. However, the thermal behavior gets complicated as the design places memory and core closer, influencing each other’s temperature. In Figure~\ref{fig:SysMemArch}(d), the core and memory are stacked, achieving the lowest interconnect delay. 
Designers prefer to place the core nearer to the heat sink for better cooling.
We refer to such a core-memory configuration as \textit{3D-stacked} in this work. \comet supports all these four core-memory configurations with various options to configure the number of cores, memory banks, and layers.
\revision{\comet also models the power dissipation from the logic core layer in the \textit{3D-ext} and \textit{2.5D} configurations.}
We perform a detailed analysis of thermal patterns for these four core-memory configurations and present the corresponding observations in Section~\ref{sec:ExpStudy}.
We built \comet to consider certain aspects of various recent emerging memory technologies where a Non-Volatile Memory~(NVM)~\cite{salkhordeh2016operating}, such as Phase Change Memory~(PCM), acts as the main memory. 
Unlike conventional DRAMs, the energy consumption for the read and write operations in PCM is considerably different. 
Hence, \comet needs to account reads and writes separately. 
\comet allows the user to specify separate parameters for the read and write energy per access, thereby providing hooks for being extended to support such emerging memory technologies. 
\revision{
However, one limitation in replacing DRAM with NVM within \comet is the underlying architectural simulator ({\em Sniper}), which does not accurately model heterogeneous read and write access times for memory. We plan to work in the future to overcome this limitation.}


\subsection{Leakage-Aware Thermal Simulation for Memories}
\label{sec:leakage-aware}
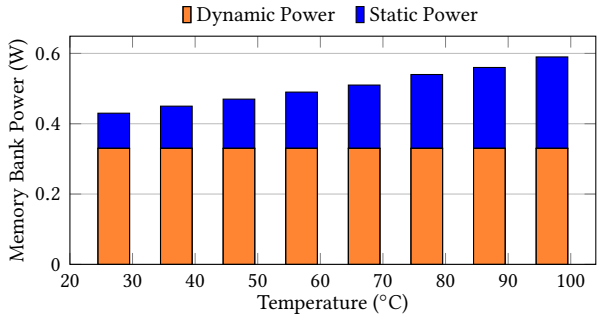
\begin{figure}[t]
\centering
    \centering




    \pgfplotsset{
        dynamic/.style={
            mem_bar,
        },
        static/.style={
            core_bar,
        },
    }
    \begin{tikzpicture}
        \begin{axis}[
			ybar stacked,
			bar width=12pt,
			ybar legend,
            width=7cm,
            height=3cm,
            ymajorgrids,
            ymin=0,
            legend columns=3,
            legend style={
                at={(0.5,1)},
                anchor=south,
            },
            xlabel={Temperature ($^\circ$C)},
            ylabel={Memory Bank Power (W)}]

            \plot[dynamic] table[x=temperature,y=dynamic,col sep=comma] {data/power_temperature.csv}; \addlegendentry{Dynamic Power}
            \plot[static] table[x=temperature,y=static,col sep=comma] {data/power_temperature.csv}; \addlegendentry{Static Power}
        \end{axis}
    \end{tikzpicture}
    \caption{Memory power dissipation versus temperature (assuming activity factor = 1 for dynamic power) 
    }
    \label{fig:Leak-Impact}
\end{figure}

As the temperature rises, the leakage power consumption increases. This increase raises the temperature, forming a temperature-leakage positive feedback loop. We model the thermal effects of temperature-dependent leakage power (for memories) similar to~\cite{siddhu2019predictncool} and validate them using \textit{ANSYS Icepak}~\cite{icepak}, a commercial detailed temperature simulator. We use {\em CACTI-3DD}~\cite{cacti} to note variations in the leakage power dissipation of the memory bank with temperature. We observe that, for memories, the leakage power contributes significantly ($\sim$40\%) to the total power dissipation (at $\sim$70\textdegree C, see Figure~\ref{fig:Leak-Impact}). In \comet, we obtain \revision{(using exponential curve fitting)} and add the temperature-dependent leakage power consumption during thermal simulation. Following a similar approach, we use {\em McPAT} to extend {\em HotSpot} to account for temperature-dependent leakage power for cores.

\subsection{Simulation Control Options} 
\label{sec:SimulationControl}

\begin{figure*}[t]
\centering
    \begin{minipage}[t]{\linewidth}
      \centering
      \includegraphics[width=\linewidth]{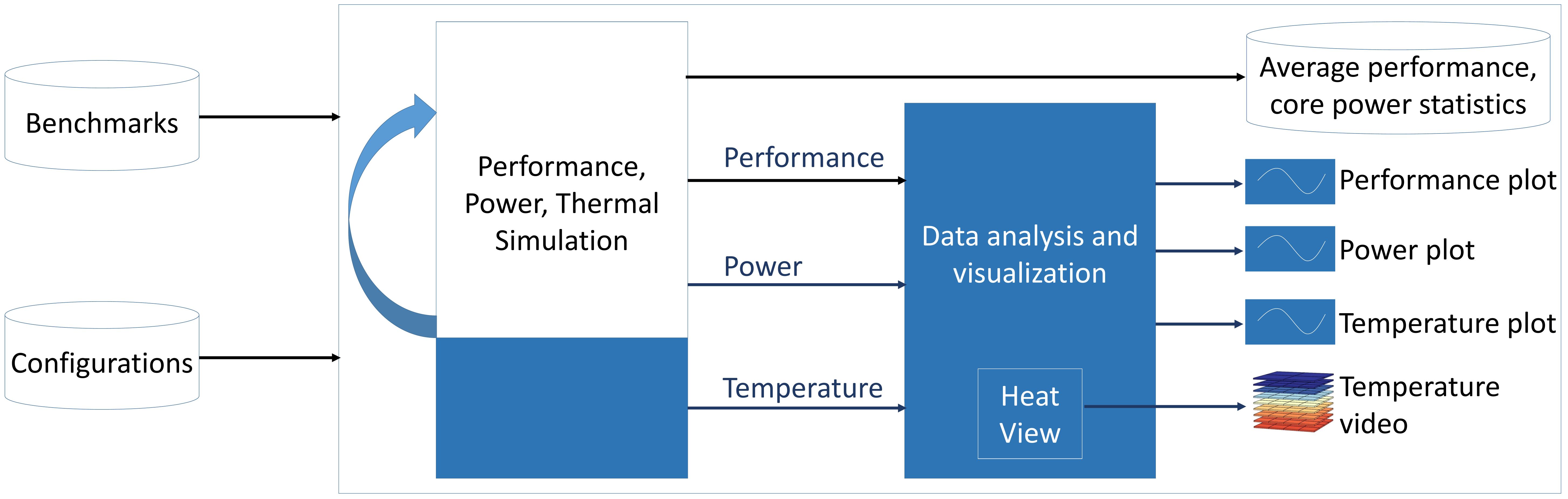}
      \caption{\SimCtrl}
      \label{fig:SimulationControl}
    \end{minipage}
\end{figure*}

A common use-case with multi-/many-core simulators is to run many simulations that vary only in a few parameters, such as the workload and architectural parameters. These simulation runs are then quantitatively compared.
\comet's \SimCtrl package provides features to facilitate this use case (Figure~\ref{fig:SimulationControl}).
It enables running simulations in batch mode and stores the traces in separate folders. 
The \SimCtrl package provides a simple Python API to specify the parameters of each simulation run: workload and \comet configuration options.
After each run, \comet stores the generated traces in a separate folder and creates plots (images) for the major metrics (power, temperature, CPU frequency, IPS, CPI stacks, etc.). 
Optionally, it also automatically creates the thermal video using the \hv feature (Section~\ref{sec:HeatView}).

In addition to  an API to run many different simulations, the \SimCtrl package provides a Python API to read the generated traces and higher-level metrics (e.g., average response time, peak temperature, and energy).
This API enables to build custom evaluation scripts. 
The \SimCtrl package, for example, can run the same random workload at varying task arrival rates with different thermal management policies. 
It generates graphs that enable users to check the resulting temperature traces visually.
Users can further perform evaluations using the \SimCtrl API~(e.g., print a table with the peak temperature of each run).

\subsection{SchedAPI: Resource Management Policies for Application Scheduling, Mapping, and DVFS} \label{sec:Scheduler}

Run-time thermal management affects the performance, power, and temperature of a multi-/many-core processor~\cite{smartboost}.
Conversely, the design of run-time thermal management techniques depends on the objective (e.g., performance or energy), the constraints (e.g., temperature), and also on the targeted platform and its characteristics (e.g., micro-architecture or cooling system). 
Thus, in research exists several thermal management techniques catering to different scenarios.
The purpose of \comet is to facilitate the development and evaluation of novel run-time thermal management techniques targeting, but not limited to, the new (stacked) core-memory configurations. 
Thermal management utilizes knobs like application scheduling, mapping and migration, and Dynamic Voltage and Frequency Scaling (DVFS). 
It then makes decisions on these knobs using observations of the system state: applications and their characteristics, power consumption, core/memory temperature, etc.
One needs to tightly integrate all these properties into the infrastructure to provide these metrics to a thermal management policy during the simulation.

Thermal management generally targets an open system, where applications arrive at times that are unknown beforehand~\cite{open_system}.
{\em HotSniper}~\cite{pathania2018hotsniper} was the first toolchain to explore the concept of scheduling for open systems using {\em Sniper}.
The scheduler API (\emph{schedAPI}) in \comet extends this feature but with a strong focus on user-friendliness to integrate new policies for mapping, migration, and DVFS.
The arrival times of the applications are configurable in several ways.
\comet supports uniform arrival times, random arrival times (Poisson distribution), or explicit user-defined arrival times.
Task mapping and migration follow the one-thread-per-core model, common in many-core processors~\cite{one_thread_per_core}.
The default policy assigns cores to an incoming application based on a static priority list.
It is straightforward to extend the default policy to implement more sophisticated policies.
DVFS uses freely-configurable voltage/frequency (V/f) levels.
\comet comes with two reference policies: a policy that assigns static frequency levels (intended to characterize the system with different applications at different V/f levels) and the Linux \textit{ondemand} governor~\cite{pallipadi2006ondemand}.
Users can configure the epoch durations (default to \SI{1}{\milli\second}) for scheduling, migration, and DVFS.
A common use-case of \comet is the implementation and study of custom resource management policies.
To this end, \emph{schedAPI} provides APIs as abstract {\em C++} classes that users can extend with custom policies, with minimal changes in the codebase.
We discuss a case study of using such a policy in \comet and corresponding insights for a stacked architecture in Section~\ref{sec:CaseStudy}.

\subsection{\hv} 
\label{sec:HeatView}
A workload executing on a typical multi-/many-core processor undergoes multiple heating and cooling phases. Such phases might occur due to the workload characteristics themselves or the effect of a DTM policy.
In such cases, developing deeper insights into the workload behavior and operation of DTM policy is essential. 
However, analyzing various log files can be cumbersome and error-prone. 
We develop and integrate \hv within \comet to analyze such thermal behavior. 
\hv generates a video to visually present the simulation's thermal behavior, with the temperature indicated through a color map. \hv takes the temperature trace file generated from \comet as input and other configurable options and generates images corresponding to each epoch and a video of the entire simulation. 
The users can use the videos corresponding to different workloads (or core-memory configurations) for comparing heating patterns across workloads (or architectures).


\hv configures according to the core-memory configurations to generate patterns.  Depending upon the specified core-memory configuration type among the four choices (\textit{2D-ext, \textit{3D-ext}, 2.5D, or 3D-stacked}), \hv can represent a core and memory stacked over each other or side-by-side. The temperature scale used within \hv to show the thermal patterns is also configurable. Additionally, to reduce the video generation time, \hv provides an option to periodically skip frames based on a user-specified sampling rate.


\hv also allows configuring of parameters to improve viewability. 
We present a 3D view of the core and memory (stacked or side-by-side) to the user by default. 
Users can specify the core or memory layer number to plot separately as a 2D map (Figures~\ref{fig:hv_3Dmem}, ~\ref{fig:hv_2.5D}, and ~\ref{fig:hv_3D}).
Figure~\ref{fig:hv_2.5D_2D} shows users can also view each layer separately.
\hv always plots the \textit{2D-ext} architecture as a 2D view (example in Figure~\ref{fig:hv_DDR}).

\subsection{Floorplan Generator} 
\label{sec:Floorplan}

Thermal simulation of a 2D processor requires a floorplan that specifies the sizes and locations of various components (cores, caches, etc.) on the silicon die. 
It requires one floorplan per layer, and a layer configuration file specifies the layer ordering and thermal properties for a stacked processor. 
\comet comes with some built-in floorplans and layer configuration files for several different architectures and configurations, as examples. 

However, in the general case of custom simulations, it is required to create floorplans and layer configuration files according to the properties of the simulated system.
\comet comes with an optional helper tool (\textit{floorplanlib}) to generate custom floorplans.
The tool supports all the four core-memory configurations described in Figure~\ref{fig:SysMemArch}.
It supports creating regular grid-based floorplans, where cores and memory banks align in a rectangular grid.
The user only needs to specify the number and dimensions of cores, memory banks, \revision{thicknesses of core or memory layers, the distance between core and memory (for 2.5D configurations), etc.}
\revision{In addition, it is possible to provide a per-core floorplan (e.g., ALU, register file, etc.), which replicates  for each core in the generated floorplan.}
User can still provide more complex (irregular) floorplans manually to \comet.

\subsection{Automated Build Verification (Smoke Testing)} 
\label{sec:SmokeTest}

While making changes to the code base, one might inadvertently introduce errors in an already working feature in the tool. To efficiently detect such scenarios, we provide an automated test suite with \comet for verifying the entire toolchain for the correct working of its key features. We use a combination of different micro-benchmarks to develop a test suite that tests various tool features. After the test completes, we summarize the pass/failure status of test cases and error logs to help users debug the causes of failure of \comet's features. While the test-suite performs a comprehensive smoke test of all \comet features, users can control and limit the testing to only a subset of features to save time. The automated build verification tests further help users test critical functionalities of \comet when they plan to extend the toolchain by adding new test cases corresponding to the added functionalities. In addition to this, it would also facilitate debugging new thermal management policies quickly. 


\section{Experimental Studies} 
\label{sec:ExpStudy}

In this section, we discuss various experiments to demonstrate the features of \comet and discuss various insights developed through these studies. Further, we also quantify the simulation time overhead of \comet over the state-of-the-art.



\subsection{Experimental Setup} \label{sec:exptSetup}






We use a diverse set of benchmark suites -- {\em PARSEC~2.1}~\cite{parsec}, {\em SPLASH-2}~\cite{splash2}, and {\em SPEC~CPU2017}~\cite{spec2017} -- to study the performance, power, and thermal profiles for core and memory. Table~\ref{table:benchmark_list} lists the selected benchmarks from each suite. We classify the benchmarks into compute-intensive~(\textit{blackscholes, swaptions, barnes, radiosity, lu.cont, raytrace, gcc, exchange, x264, nab, mcf}), mixed~(\textit{streamcluster, vips, dedup, bodytrack, water.nsq, cholesky}), and memory-intensive~(\textit{lbm}, \textit{mcf}) based on their memory access rate.
We compile the source code for {\em PARSEC~2.1} (with input size \textit{simmedium}) and {\em SPLASH2} benchmarks (with input size \textit{small}) to get the binaries for simulation. We directly use pre-generated traces (Pinballs) from~\cite{specTraces} for simulation (with 100M instructions) for {\em SPEC~CPU2017} benchmarks.

Table~\ref{table:CoreMemParams} shows the core and memory parameters for various core-memory configurations that we use in our experiments. We use \comet's automated \textit{floorplanlib} tool to generate the floorplans for various core-memory configurations.
We run simulations using \comet and obtain performance, power, and temperature metrics for various workloads. \revision{\textit{HotSpot} uses grid-level simulation with an 8x8 grid in the center mode.}
Thermal simulation is invoked periodically at \SI{1}{\milli\second} frequency.


\begin{table}[t]
  \caption{Core and Memory Parameters}
  \centering
    \begin{tabular}{|p{3.6cm}|p{7.7cm}|}
    \hline
    \textbf{Core Parameter} & \textbf{Value}\tabularnewline
    \hline
    \hline
    Number of Cores& 4\tabularnewline
    \hline
    Core Model & 3.6 GHz, 1.2 V, 22 nm, out-of-order, 3 way decode, 84 entry ROB, 32 entry LSQ\tabularnewline
    \hline
    L1 I/D Cache & 4/16 KB, 2/8-way, 64B-block \tabularnewline
    \hline
    L2 Cache & Private, 64 KB, 8-way/64B-block\tabularnewline
    \hline
    \hline
    \textbf{Memory Parameter} & \textbf{Value}\tabularnewline
    \hline
    \hline
    3D Memory ({\em 3D-ext}, {\em 2.5D}, {\em 3D-stacked})
 Configuration & 1 GB, 8 layer, 16 channels, 8 ranks, 1 bank per rank, closed page policy, 29/20/15 ns (latency), 7.6 GBps (per channel bandwidth)\tabularnewline
    \hline
    2D Memory Off-chip Configuration & 2 GB, 1 layer, 1 channel, 4 ranks, 4 bank per rank, closed page policy, 45 ns (latency), 7.6 GBps (per channel bandwidth)\tabularnewline
    \hline    
    \end{tabular}
    \label{table:CoreMemParams}
\end{table}

\begin{table}[t]
  \caption{List of Benchmarks}
  \centering
    \begin{tabular}{|p{3.3cm}|p{8.7cm}|}
    \hline
    \textbf{Benchmark Suite} & \textbf{Selected Benchmarks}\tabularnewline
    \hline
    \hline
    {\em PARSEC~2.1} & \textit{dedup, streamcluster, vips, bodytrack, swaptions, blackscholes}\tabularnewline
    \hline
    {\em SPLASH-2} & \textit{lu.cont, water.nsq, radiosity, raytrace, barnes, cholesky}\tabularnewline
    \hline
    {\em SPEC~CPU2017} & \textit{lbm, mcf, gcc, nab, x264, exchange} \tabularnewline
    \hline
    \end{tabular}
    \label{table:benchmark_list}
\end{table}

\subsection{Thermal Profile for Various Architecture Configurations}
\label{sec:ThermalProfileArchitecture}

\begin{figure}[t]
\centering
      \includegraphics[width=0.70\linewidth]{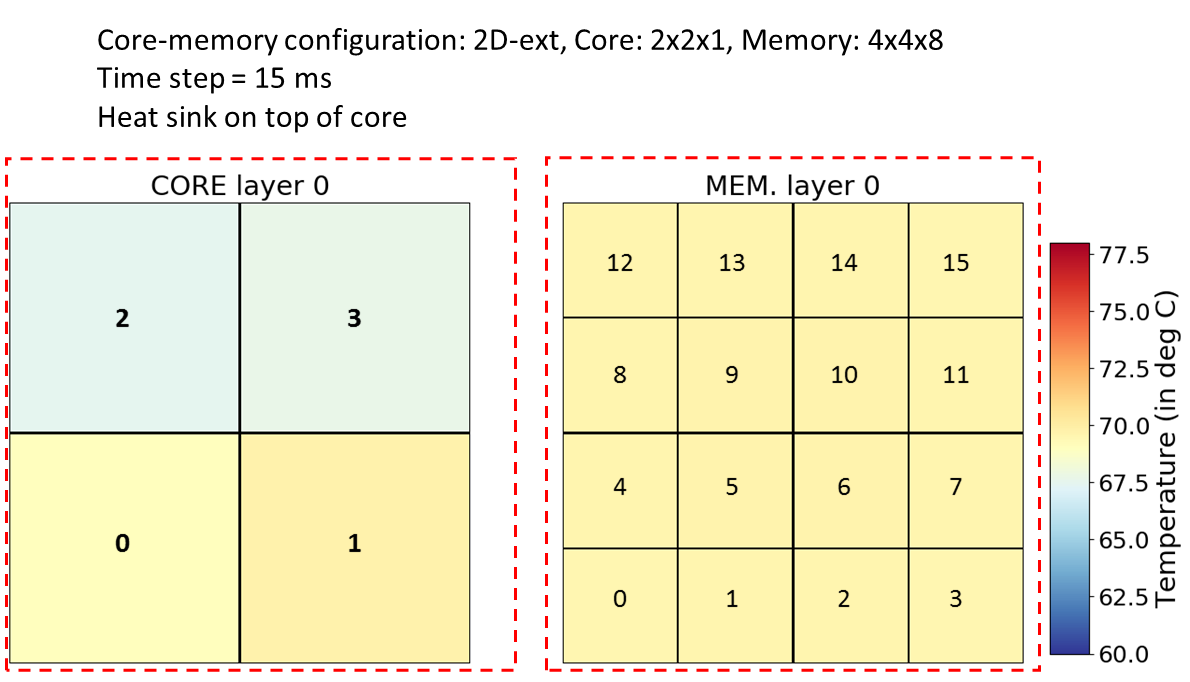}
      \caption{\revision{Thermal profile of core and memory at \SI{15}{\milli\second} when executing a heterogeneous workload on {\em 2D-ext} core-memory configuration.}}
      \label{fig:hv_DDR}
\end{figure}
\begin{figure}[t]
\centering
      \includegraphics[width=0.90\linewidth]{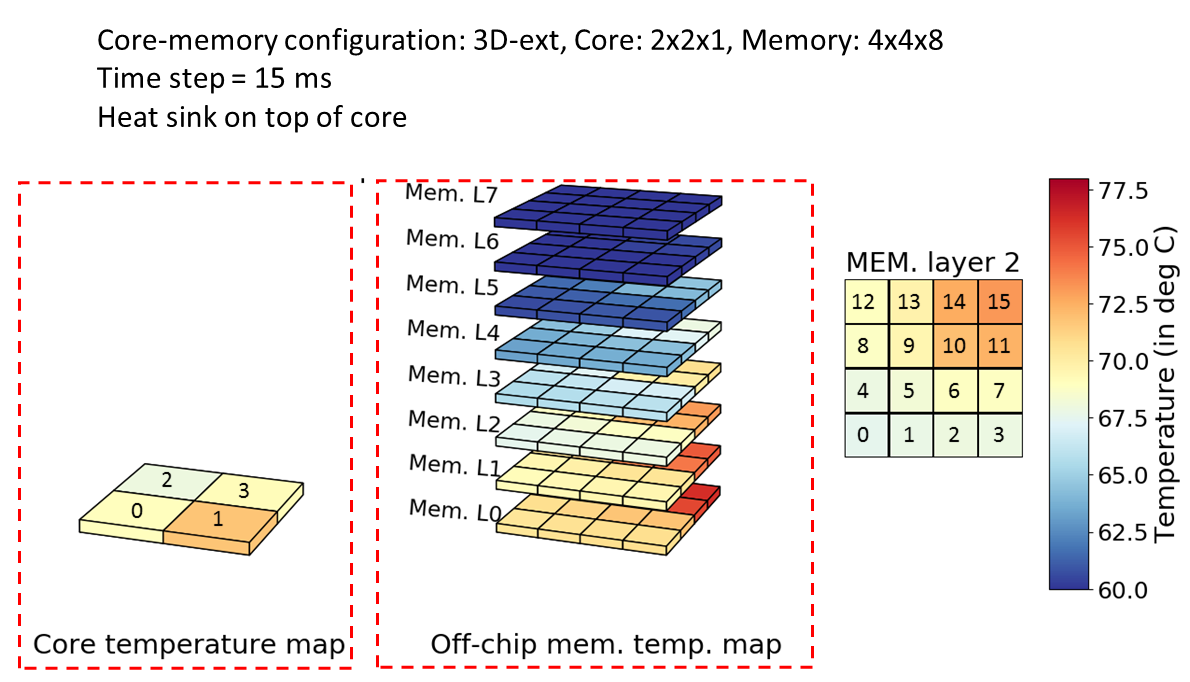}
      \caption{\revision{Thermal profile of core and memory at \SI{15}{\milli\second} when executing a heterogeneous workload on {\em 3D-ext} core-memory configuration.}}
      \label{fig:hv_3Dmem}
\end{figure}
\begin{figure}[t]
      \includegraphics[width=0.90\linewidth]{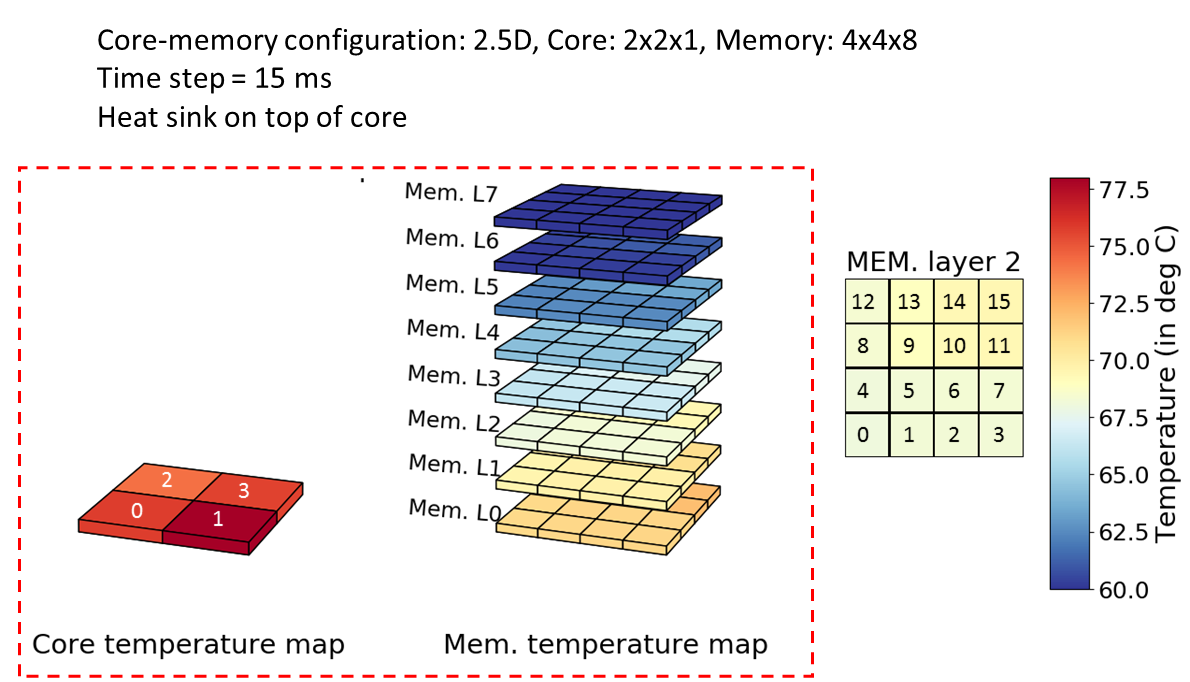}
      \caption{\revision{Thermal profile of core and memory at \SI{15}{\milli\second} when executing a heterogeneous workload on {\em 2.5D} core-memory configuration.}}
      \label{fig:hv_2.5D}
\end{figure}
\begin{figure}[t]
\centering
      \includegraphics[width=0.90\linewidth]{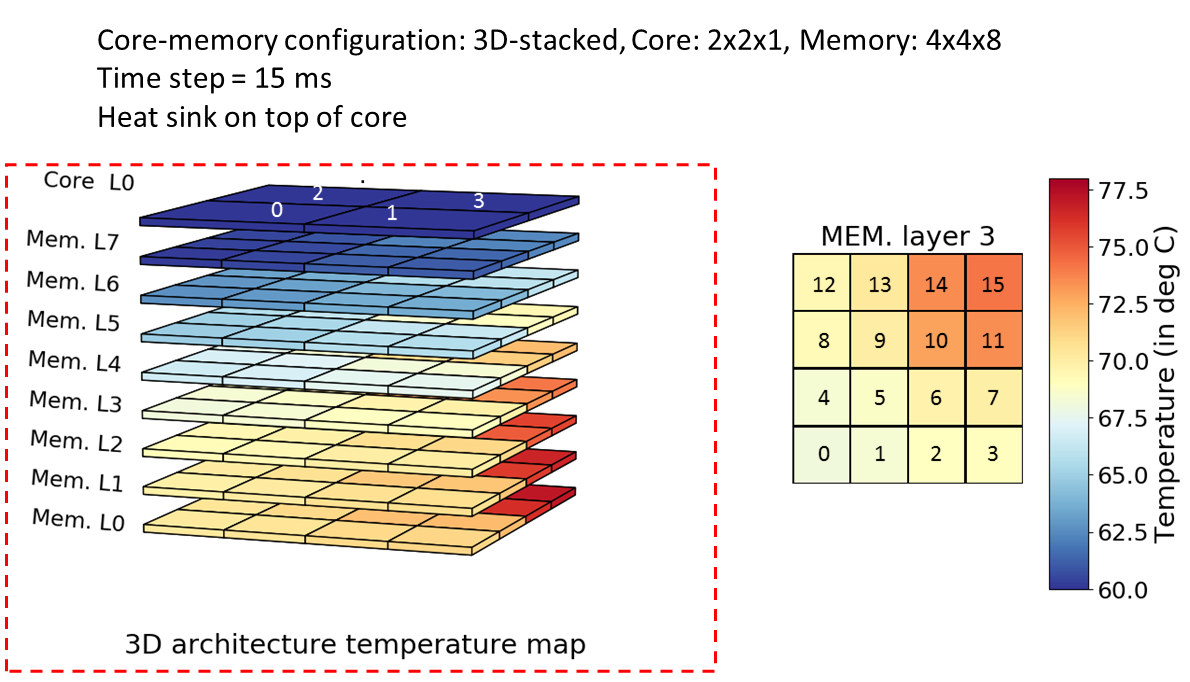}
      \caption{\revision{Thermal profile of core and memory at \SI{15}{\milli\second} when executing a heterogeneous workload on {\em 3D-stacked} core-memory configuration.}}
      \label{fig:hv_3D}
\end{figure}

\revision{
We present the thermal behavior of cores and memories for each of the four core-memory configurations supported by \comet. We consider \textit{exchange}, \textit{x264}, \textit{mcf}, and \textit{lbm} benchmarks from the {\em SPEC~CPU2017} suite and map them on Cores 0, 1, 2, and 3, respectively, to exercise a heterogeneous workload containing benchmarks of different memory intensity. Each core maps to a fixed set of 3D memory channels. Core 0 maps to Channels \{0, 1, 4, 5\}, Core 1 maps to Channels \{2, 3, 6, 7\}, Core 2 maps to Channels \{8, 9, 12, 13\}, and Core 3 maps to Channels \{10, 11, 14, 15\}.}
 \hv uses the temperature trace generated during the simulation to create a video of the thermal pattern of various cores and memory banks. 
The videos for the simulations are available online at   \href{run:https://tinyurl.com/cometVideos}{tinyurl.com/cometVideos}. Figures~\ref{fig:hv_DDR}, ~\ref{fig:hv_3Dmem}, ~\ref{fig:hv_2.5D}, and ~\ref{fig:hv_3D} present snapshots at \SI{15}{\milli\second} of simulation time for each of the four architectures.

\revision{
Figure~\ref{fig:hv_DDR} presents the temperature profile of cores and the external DDR memory. We observe that Cores 0 and 1 have relatively higher temperatures than Cores 2 and 3 due to the execution of compute-intensive benchmarks on Cores 0 and 1. Further, Core 1 has a slightly higher temperature than Core 0 as \textit{x264} is more compute-intensive than \textit{exchange}. We do not observe any temperature gradient on the memory side. We consider a single channel for the 2D memory with accesses from different cores shared and uniformly distributed among banks, thereby eliminating any gradient.
}
Figure~\ref{fig:hv_3Dmem} shows the temperature profile of cores and an external 8-layer 3D memory. 
As cores and memory banks  physically locate on different chips, they do not influence each other's temperature and have different thermal profiles.
We see that the memory banks attain significantly higher temperatures. Further, due to the heat sink at the top of the 3D memory, the temperature of the lower layers is higher than that of the upper layers, with a gradual decrease as we move up the memory stack.
\revision{
Due to the heterogeneous nature of benchmarks and each core mapping to a fixed set of channels, we observe that different 3D memory channels attain different temperatures. In the cross-section view of a memory layer shown in the figure, Channels 10, 11, 14, and 15 correspond to \textit{lbm}, a highly memory-intensive benchmark. 
\textit{lbm} is a highly memory-intensive benchmark that results in high temperatures in the memory layer. 
Channels 0, 1, 4, and 5 are relatively cooler as they correspond to Core 0, which executes a compute-intensive benchmark (\textit{exchange}). Channels 2, 3, 6, and 7 also correspond to a compute-intensive benchmark (\textit{x264}), but Channels 6 and 7 have higher temperatures than Channels 2 and 3 due to thermal coupling from adjacent hot Channels 10 and 11.
Different cores also attain different temperatures due to the differing nature of the benchmarks executed.
}
While this core-memory configuration (\textit{3D-ext}) differs from \textit{2D-ext} only in terms of using an external 3D memory compared to a DDR memory, we observe that the cores in \textit{3D-ext} (Figure~\ref{fig:hv_3Dmem}) are relatively hotter than the cores in the \textit{2D-ext} (Figure~\ref{fig:hv_DDR}) because of faster execution enabled by the 3D memory. 
\comet enables such insights due to the integrated core-memory thermal simulation that cannot be easily quantified (accurately) when using a standalone trace-based simulation infrastructure.

\begin{figure}[t]
      \centering
      \includegraphics[width=0.90\linewidth]{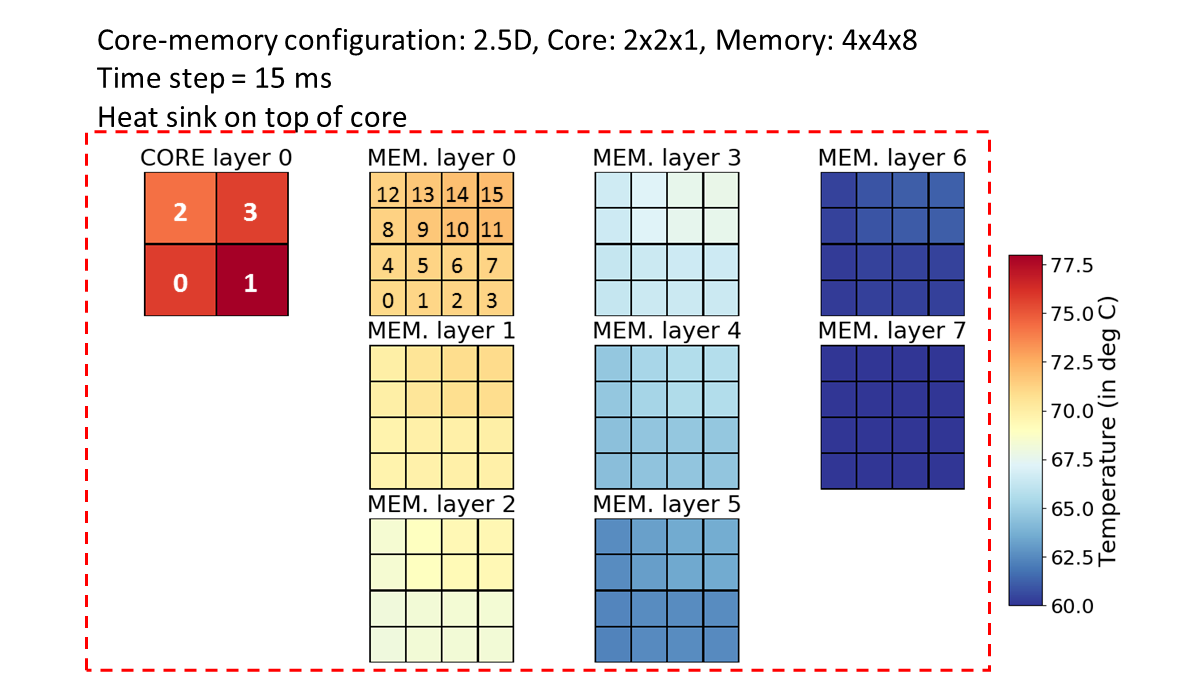}
      \caption{\revision{Detailed/2D view of each layer for the {\em 2.5D} configuration, corresponding to Figure~\ref{fig:hv_2.5D}}}
      \label{fig:hv_2.5D_2D}
\end{figure}

Figure~\ref{fig:hv_2.5D} shows the temperature profile of cores and 3D memory integrated on the same package in a {\em 2.5D} configuration. 
\revision{
Similar to the previous case of {\em 3D-ext} (Figure~\ref{fig:hv_3Dmem}), we observe that different cores and different memory channels in the {\em 2.5D} core-memory configuration attain different temperatures due to the heterogeneous nature of the workload.
Also, the core and memory are thermally coupled in the {\em 2.5D} core-memory configuration, resulting in significantly higher temperatures for the same workload. Since the cores are away from the heat sink compared to {\em 3D-ext}, their heat dissipation capability reduces, leading to much higher temperatures.
}

Figure~\ref{fig:hv_3D} shows the thermal profile for a {\em 3D-stacked} configuration with one layer of four cores stacked over an 8-layer 3D memory. We observe that any layer of the memory is hotter than the corresponding layer in the \textit{3D-ext} or {\em 2.5D} core-memory configuration due to the increased stacking of cores on top of the 3D memory, limiting the heat dissipation paths further raising the temperature. 
\revision{
Similar to other core-memory configurations, we observe that different memory channels attain different temperatures due to the heterogeneous nature of the workload. However, the cores heat almost uniformly given their proximity to the heat sink and their coupling with the memory layers. While Cores 0 and 1 executing compute-intensive benchmarks should attain higher temperatures than Cores 2 and 3, their corresponding memory channels exhibit lower temperatures due to fewer memory accesses and help absorb the excess heat. \comet enables such insights due to its support for various core-memory configurations.
}

To illustrate the feature of \hv to create thermal maps with detailed layerwise details (2D view), we use the {\em 2.5D} configuration (Figure~\ref{fig:hv_2.5D}) as an example. The corresponding layerwise plot is shown in Figure~\ref{fig:hv_2.5D_2D} and provides more details of each layer.

\subsection{Thermal Profile for Various Benchmarks}
\label{sec:ThermalProfileBenchmarks}

\begin{figure*}[t]
    \centering
    \pgfplotstableread[col sep=comma]{data/benchmark_characterization.csv}\datatable
    \pgfplotsset{
        benchmark_characterization_axis/.style={
			ybar=0pt,
			bar width=4pt,
			ybar legend,
            width=11.5cm,
            height=4.5cm,
            xtick=data,
            enlarge x limits={true,abs value=0.6},
			major tick length=0,
            xticklabels from table={\datatable}{application},
            xticklabel style={
                font=\em,
                rotate=30,
                anchor=north east,
                yshift=0mm,
                xshift=0mm,
            },
            legend columns=3,
            legend style={
                at={(0.5,1)},
                anchor=south,
            },
        },
        core_temp/.style={
            ybar,
            bar shift={-\pgfplotbarwidth},
            core_bar,
            draw=black,
        },
        mem_temp/.style={
            ybar,
            bar shift={0},
            mem_bar,
            draw=black,
        },
        exec_time/.style={
            ybar,
            bar shift={\pgfplotbarwidth},
            other_bar,
            draw=black,
        },
    }
    \begin{tikzpicture}
        \begin{axis}[
            benchmark_characterization_axis,
            axis y line*=left,
            ymajorgrids,
            ymin=65,
            ymax=80,
            ytick distance=5,
            xlabel={},
            ylabel={Core / Memory Temperature ($^\circ$C)}]

            \plot[core_temp] table[x expr=\coordindex,y=core_temp] {\datatable};
            \plot[mem_temp] table[x expr=\coordindex,y=mem_temp] {\datatable};
            \addlegendimage{exec_time}
            \legend{Core Temperature,Memory Temperature,Execution Time}
        \end{axis}
        \begin{axis}[
            benchmark_characterization_axis,
            axis y line*=right,
            axis x line=none,
            ymin=0,
            ymax=1000,
            ytick distance=250,
            xlabel={},
            ylabel={Execution Time (ms)},
	        ylabel style={yshift=2.5mm}]

            \plot[exec_time] table[x expr=\coordindex,y=exec_time] {\datatable};
        \end{axis}
    \end{tikzpicture}

    \caption{Temperature for three different benchmark suites running on 4 cores and off-chip 3D-DRAM memory architecture: PARSEC (with {\em simmedium} input size), SPLASH2 (with {\em small} input size), and SPEC CPU2017 (with 100 million instructions)}
    \label{fig:tss}
\end{figure*}
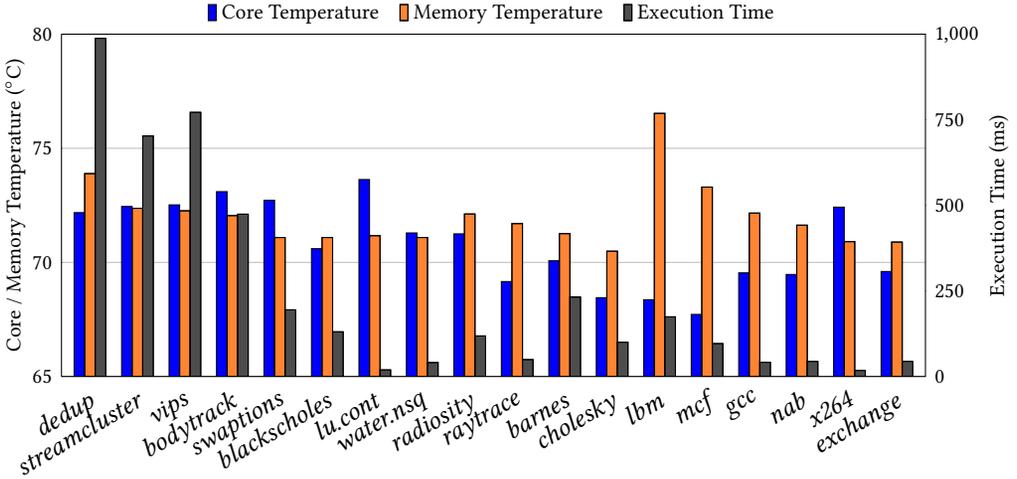

We analyze the performance, power, and thermal profile for core and memory for various benchmark suites using \comet. Figure~\ref{fig:tss} shows the core, memory temperature, and execution time for {\em PARSEC~2.1}, {\em SPLASH-2}, and {\em SPEC~CPU2017} benchmarks running on a four-core system with an off-chip 3D memory (\textit{3D-ext} architecture). 
\revision{In these experiments, we execute multiple instances of the  benchmark, one on each CPU core.} A four-core system with a heat sink has sufficient cooling paths. However, we see a significant temperature rise with higher power dissipation in cores.

Most benchmarks in {\em PARSEC~2.1} and {\em SPLASH-2} suite are multi-threaded and compute-intensive. So the average DRAM access rate remains low for these benchmarks throughout their execution. However, due to the high density in 3D memories, the leakage power (temperature-dependent) dissipation contributes significantly to overall memory power dissipation. Stacking also increases the power density, resulting in memory temperatures of around 71$^\circ$C (increasing with memory access rate). For the {\em SPEC~CPU2017} suite, \textit{lbm} is a memory-intensive benchmark with a high average DRAM access rate (as high as $10^8$ accesses per second).
Therefore, \textit{lbm} results in significantly higher memory temperatures than other benchmarks.

\begin{figure}[t]
      \centering
      \includegraphics[width=0.99\linewidth]{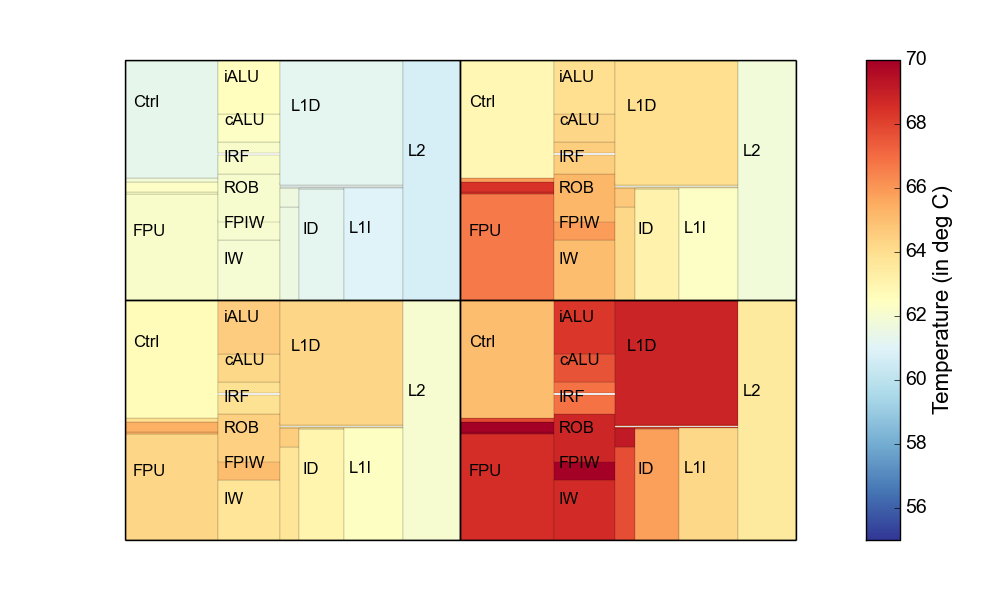}
      \caption{\revision{Thermal profile of cores with a fine-grained floorplan in a \emph{3D-ext} core-memory configuration}}
      \label{fig:subcore}
\end{figure}

\revision{\subsection{Thermal Profile with Fine-grained Core Components}}
\label{sec:subcore}
\revision{
We illustrate the ability of \comet to simulate a fine-grained core floorplan with individual components of a core modeled explicitly. As mentioned in Section~\ref{sec:Floorplan}, \textit{floorplanlib} can generate a multi-core floorplan if the user also provides a fine-grained floorplan for a single core as input. We obtain the area of each component from \textit{McPAT} to obtain the floorplan for a single core. The relative placement of different components is similar to {\em Intel's Skylake} processor design from
HotGauge~\cite{Hankin:2021:Hotgauge}.
\textit{floorplanlib} generates a fine-grained floorplan for four cores using this single-core floorplan as a template. We use the four-core floorplan to simulate workloads in a \textit{3D-ext} configuration.
We use the same workloads and 3D memory configuration as in our previous experiments in Section~\ref{sec:exptSetup} and a finer grid size of 32x32. Figure~\ref{fig:subcore} shows the corresponding thermal map obtained from \comet.\footnote{\revision {The thermal map figure is generated outside of \hv as currently \hv supports plotting of uniform blocks only.}} 
Similar to our previous result for the \textit{3D-ext} configuration shown in Figure~\ref{fig:hv_3Dmem}, we observe that different cores attain different temperatures due to heterogeneous workloads. In addition, due to consideration of a fine-grained floorplan with their power consumption, we observe the presence of a thermal gradient between different components of the same core. 
We observe that the execution units like the ALU (Arithmetic and logic unit), FPU (Floating point processing unit), ROB (Reorder buffer), etc., attain a higher temperature than the rest of the components such as ID (Instruction decoder), L1I (L1 instruction cache), or the L2 cache. 
Such a feature present in \comet can provide deeper insights about thermal hotspots within a core. Accordingly, one can take more appropriate thermal management decisions.
}

\revision{\subsection{Effect of thermal coupling in 2.5D architecture}}

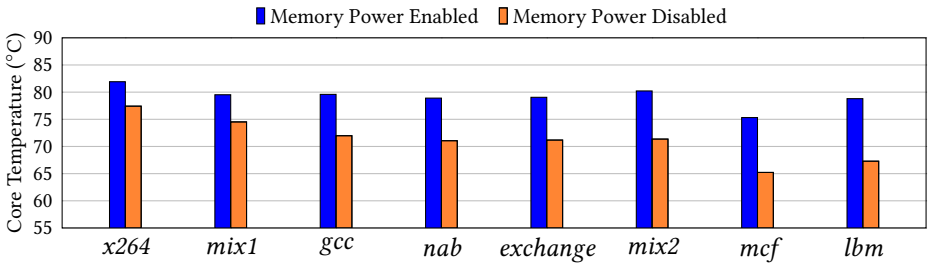
\begin{figure*}[t]
    \centering
    \pgfplotstableread[col sep=comma]{data/2.5D_coupling.csv}\datatable
    \pgfplotsset{
        2.5D_coupling_axis/.style={
			ybar=0pt,
			bar width=6pt,
			ybar legend,
            width=11.5cm,
            height=2.5cm,
            xtick=data,
            enlarge x limits={true,abs value=0.6},
			major tick length=0,
            xticklabels from table={\datatable}{application},
            xticklabel style={
                font=\em,
                yshift=0mm,
                xshift=0mm,
            },
            legend columns=3,
            legend style={
                at={(0.5,1)},
                anchor=south,
            },
        },
        memory_enabled/.style={
            ybar,
            core_bar,
            draw=black,
        },
        memory_disabled/.style={
            ybar,
            mem_bar,
            draw=black,
        },
    }
    \begin{tikzpicture}
        \begin{axis}[
            2.5D_coupling_axis,
            ymajorgrids,
            ymin=55,
            ymax=90,
            ytick distance=5,
            xlabel={},
            ylabel={Core Temperature ($^\circ$C)}]

            \plot[memory_enabled] table[x expr=\coordindex,y=memory_enabled] {\datatable};
            \plot[memory_disabled] table[x expr=\coordindex,y=memory_disabled] {\datatable};
            \legend{Memory Power Enabled, Memory Power Disabled}
        \end{axis}
       
    \end{tikzpicture}

    \caption{\revision{Core temperature when 3D memory power modeling is enabled and disabled to show thermal coupling in \textit{2.5D} core-memory configuration}}
    \label{fig:coupling}
\end{figure*}


\revision{We illustrate the effect of thermal coupling between the core and memory in a \textit{2.5D} core-memory configuration. As the 3D memory co-locates with the cores on the same package, memory temperature affects the core temperature and vice-versa. 
We experiment with the 3D memory power enabled (both leakage and dynamic power taken into account) and 3D memory power disabled (leakage and dynamic power forced to 0) during simulation.
We repeat this experiment for eight different workloads to exercise different activity factors for cores and memory. 
We use six homogeneous workloads (Table~\ref{table:benchmark_list}) as used in previous experiments and two heterogeneous workloads, with each workload consisting of four independent benchmarks. The \textit{mix1} heterogeneous workload includes a mix of \textit{lbm}, \textit{x264}, \textit{exchange}, and \textit{mcf} benchmarks, while the \textit{mix2} workload includes \textit{lbm}, \textit{gcc}, \textit{nab}, and \textit{mcf} benchmarks.
Figure~\ref{fig:coupling} shows the maximum core temperature (out of the four cores) for memory power being enabled and disabled, with workloads ordered as per increasing memory intensity. We observe that the thermal coupling increases as we move from compute-intensive workloads (towards left) to memory-intensive workloads (towards the right).
A higher memory activity raises the memory temperature, leading to higher thermal coupling. Memory-intensive workloads (e.g., \textit{lbm}, \textit{mcf}) induce maximum thermal coupling, and enabling 3D memory power dissipation raises the temperature of cores by up to 11 \textdegree C (for \textit{lbm}). 
}

\subsection{Case Study: Thermal-Aware Scheduler and DVFS} \label{sec:CaseStudy}

We show in this section a case study of the analyses that are possible with \comet and demonstrate some trends that appear in stacked core-memory configurations.
We employ the Linux \emph{ondemand} governor~\cite{pallipadi2006ondemand} with DTM.
The \emph{ondemand} governor increases or decreases per-core V/f-levels when the core utilization is high or low, respectively.
DTM throttles the chip to the minimum V/f-level when some thermal threshold exceeds  and increases the frequency back to the previous level if the temperature falls below the thermal threshold minus a hysteresis parameter.
In this experiment, we set the two thresholds to 80$^\circ$C and 78$^\circ$C.
The temperature is initialized to a 70$^\circ$C peak to emulate a prior workload execution.

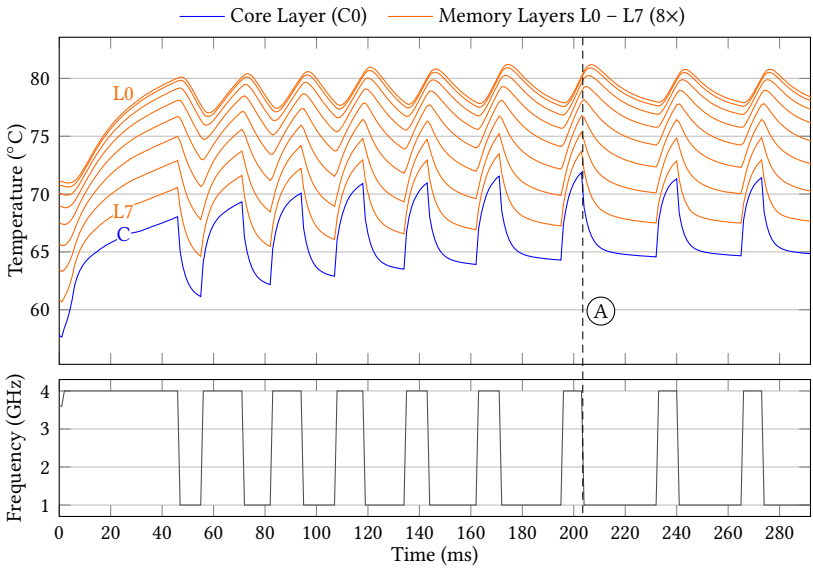
\begin{figure}
    \centering
    \pgfplotsset{
        case_study_axis/.style={
            width=10.0cm,
            xmin=0,
            xmax=292,
            xlabel={Time (ms)},
            ymajorgrids,
            legend columns=3,
            legend style={
                at={(0.5,1)},
                anchor=south,
            },
        },
        freq/.style={
            other_color,
            fill=none,
        },
        core/.style={
            core_color,
            fill=none,
        },
        memory/.style={
            mem_color,
            fill=none,
        }
    }


        \begin{tikzpicture}
            \begin{axis}[
                    case_study_axis,
                    name=temp_axis,
                    height=4.3cm,
                    xlabel={},
                    xticklabels={,,},
                    ylabel={Temperature ($^\circ$C)},
        	        ylabel style={yshift=-0.5mm},
                    ]
                \addlegendimage{core}
                \addlegendimage{memory}
                \legend{Core Layer (C0),Memory Layers L0 -- L7 ($8\times$)}

                \plot[core] table[x=t,y=core_peak_temp,col sep=comma] {data/case_study.csv};
                \plot[memory] table[x=t,y=mem_peak_temp_0,col sep=comma] {data/case_study.csv};
                \plot[memory] table[x=t,y=mem_peak_temp_1,col sep=comma] {data/case_study.csv};
                \plot[memory] table[x=t,y=mem_peak_temp_2,col sep=comma] {data/case_study.csv};
                \plot[memory] table[x=t,y=mem_peak_temp_3,col sep=comma] {data/case_study.csv};
                \plot[memory] table[x=t,y=mem_peak_temp_4,col sep=comma] {data/case_study.csv};
                \plot[memory] table[x=t,y=mem_peak_temp_5,col sep=comma] {data/case_study.csv};
                \plot[memory] table[x=t,y=mem_peak_temp_6,col sep=comma] {data/case_study.csv};
                \plot[memory] table[x=t,y=mem_peak_temp_7,col sep=comma] {data/case_study.csv};

                \node[font=\footnotesize,mem_color,fill=white,inner sep=0pt] at (axis cs:25,78.7) {L0};
                \node[font=\footnotesize,mem_color,fill=white,inner sep=0pt] at (axis cs:25,68.5) {L7};
                \node[font=\footnotesize,core_color,fill=white,inner sep=0pt] at (axis cs:25,66.5) {C};
            \end{axis}
            \begin{axis}[
                    case_study_axis,
                    name=freq_axis,
                    at=(temp_axis.south),
                    yshift=-2mm,
                    anchor=north,
                    height=1.8cm,
                    ylabel={Frequency (GHz)},
        	        ylabel style={yshift=1mm},
        	        ytick={1,2,3,4}
                    ]
                \plot[freq] table[x=t,y=peak_frequency,col sep=comma] {data/case_study.csv};
                \coordinate (poi) at (axis cs:203.5,0);
            \end{axis}
            \draw[densely dashed] (poi |- temp_axis.north) to  (poi |- freq_axis.south);
            \node[font=\footnotesize,circle,draw,fill=white,anchor=west,inner sep=1pt] at ([xshift=0.5mm,yshift=0.7cm]poi |- temp_axis.south) {A};
        \end{tikzpicture}
    \caption{
        Transient temperature of the hotspot in each of the nine layers of a 3D architecture (one core layer, eight memory layers).
        The memory layer farthest from the heatsink forms the overall system hotspot, and determines when DTM throttles the system.
    }
    \label{fig:case_study}
\end{figure}

We execute the \emph{PARSEC} \emph{swaptions} with four threads to fully utilize all processor cores.
Figure~\ref{fig:case_study} depicts the temperature and frequency throughout the execution.
\emph{Swaptions} is compute-bound, and hence the \emph{ondemand} governor selects the highest available frequency.
Consequently, the processor reaches the temperature limit of 80$^\circ$C fast. DTM then reduces the frequency until the temperature has decreased, leading to thermal cycles as shown in the figure, where the frequency toggles between a low and a high value.
We observe the peak temperature not on the cores but the memory layers (Section~\ref{sec:ThermalProfileArchitecture}).

This simulation uses a {\em 3D-stacked} architecture---enabled by \comet---and shows some interesting trends.
The temperature on the core layer is directly affected by DTM.
The temperature immediately reduces almost exponentially upon thermal throttling, e.g., at 203\,ms (Point~A).
It takes several milliseconds until the temperature at layers farther away from the core layer reduces (in this example 5\,ms), during which the temperature overshoots the thermal threshold.
Similarly, when returning to normal operation, the temperature of the hotspot reacts with a significant delay to DTM decisions.
This delay is because the hotspot's location (lowest memory layer) is far from the layer most affected by DTM (core layers).
This observation is unlike traditional 2D architectures, where the two coincide (thermal hotspots in the cores).
Existing state-of-the-art thermal management~\cite{rapp2020neural} and power budgeting algorithms~\cite{niknam2021t} cannot account for these trends.  
Therefore, such different trends require novel policies (algorithms) that can be easily evaluated on \comet using the interfaces presented in Section~\ref{sec:Scheduler}. 

\subsection{Parameter Variation}
\label{sec:Parameter}

\subsubsection{Increasing the Number of Cores}
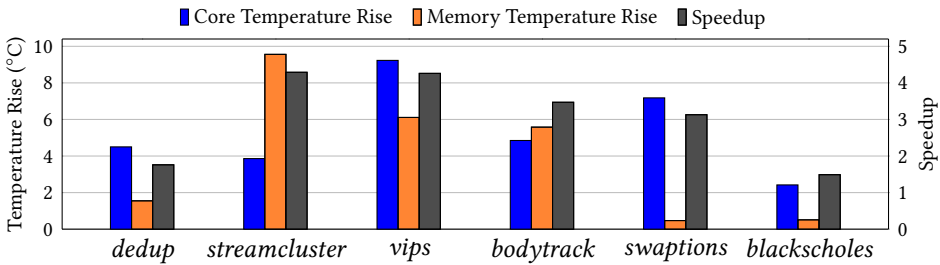
\begin{figure}[t]
\centering
    \centering
    \pgfplotstableread[col sep=comma]{data/arch_comparison.csv}\datatable
    \pgfplotsset{
        arch_comparison_axis/.style={
			ybar=0pt,
			bar width=8pt,
			ybar legend,
            width=11cm,
            height=2.5cm,
            xtick=data,
            enlarge x limits={true,abs value=0.6},
			major tick length=0,
            xticklabels from table={\datatable}{application},
            xticklabel style={font=\em},
            legend columns=3,
            legend style={
                at={(0.5,1)},
                anchor=south,
            },
        },
        core_temp_rise/.style={
            ybar,
            bar shift={-\pgfplotbarwidth},
            core_bar,
            draw=black,
        },
        mem_temp_rise/.style={
            ybar,
            bar shift={0},
            mem_bar,
            draw=black,
        },
        speedup/.style={
            ybar,
            bar shift={\pgfplotbarwidth},
            other_bar,
            draw=black,
        },
    }
    \begin{tikzpicture}
        \begin{axis}[
            arch_comparison_axis,
            axis y line*=left,
            ymajorgrids,
            ymin=0,
            ymax=10.4,
            ytick distance=2,
            xlabel={},
            ylabel={Temperature Rise ($^\circ$C)}]

            \plot[core_temp_rise] table[x expr=\coordindex,y=core_temp_rise] {\datatable};
            \plot[mem_temp_rise] table[x expr=\coordindex,y=mem_temp_rise] {\datatable};
            \addlegendimage{speedup}
            \legend{Core Temperature Rise,Memory Temperature Rise,Speedup}
        \end{axis}
        \begin{axis}[
            arch_comparison_axis,
            axis y line*=right,
            axis x line=none,
            ymin=0,
            ymax=5.2,
            ytick distance=1,
            xlabel={},
            ylabel={Speedup},
	        ylabel style={yshift=2.5mm}]

            \plot[speedup] table[x expr=\coordindex,y=speedup] {\datatable};
        \end{axis}
    \end{tikzpicture}
    \caption{
        Speed-up and steady-state temperature rise for 16-core configuration (normalized to 4-core)
    }
    \label{fig:Parsec-16C}
\end{figure}

We study the performance and thermal effect of increasing the number of cores (and threads) for the {\em PARSEC} workloads running on {\em 3D-ext} configuration. We increase the number of cores from 4 to 16 and observe that some {\em PARSEC} workloads~(such as \textit{bodytrack}, \textit{streamcluster}, \textit{vips}, and \textit{swaptions}) can utilize parallelism more effectively (Figure~\ref{fig:Parsec-16C}). Workloads such as \textit{blackscholes} and \textit{dedup} either have a significant serial phase or imbalanced thread execution time, resulting in a limited speedup with a marginal increase in temperature. 
For \textit{blackscholes}, we observe a speedup of $\sim$1.5x (compared to a 4x increase in the number of cores) as it spends most of the execution time in the serial phase.

\subsubsection{Increasing the Number of Core Layers}
\begin{figure}[t]
      \centering
      \includegraphics[width=0.90\linewidth]{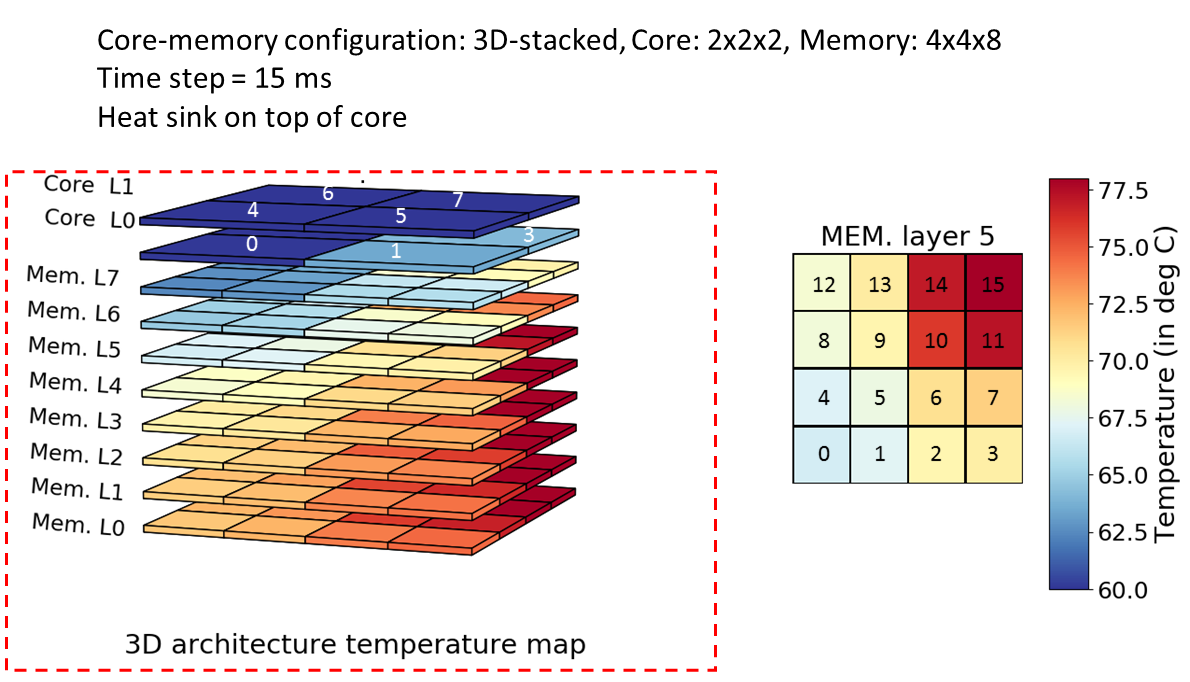}
      \caption{\revision{Thermal profile of core and memory at \SI{15}{\milli\second} when executing a heterogeneous workload on \textit{3D-stacked} core-memory configuration with 2 layers of core on top of 8 layers of memory.}}
      \label{fig:hv_3D_2L}
\end{figure}

Until now, all our experiments have considered cores on a single layer. Here, we demonstrate the ability of \comet to perform thermal simulation for multiple layers of cores. We consider the same \textit{3D-stacked} core-memory configuration corresponding to Figure~\ref{fig:hv_3D} but extend it to have two layers of cores and, therefore, a total of 8 cores.
\revision{
We execute the same set of heterogeneous benchmarks, with the same benchmark mapped to vertically stacked cores. Specifically, \textit{exchange}, \textit{x264}, \textit{mcf}, and \textit{lbm} are mapped to cores \{0,4\}, \{1,5\}, \{2,6\}, and \{3,7\} respectively. Figure~\ref{fig:hv_3D_2L} shows the temperature pattern of various layers of core and memory. We observe that, compared to Figure~\ref{fig:hv_3D} with only a single core layer, an additional layer of core on top raises the temperatures of the bottom layers significantly. 
Further, the temperature gradient (effect of adjacent layers) is more pronounced with two core layers than one core layer (Figure~\ref{fig:hv_3D_2L}). 
}

This experiment demonstrates the versatility of \comet in adapting to different kinds of core-memory configurations with single/multiple layers of cores integrated with single/multiple layers of memory. Such a capability enables \comet to analyze the performance, power, and thermal behavior of various emerging core-memory configurations and identify various optimization opportunities within them. We strongly believe that \comet could help identify many newer research problems and evaluate their proposed solutions.

\subsection{Overhead Analysis}
\label{sec:Overhead Analysis}

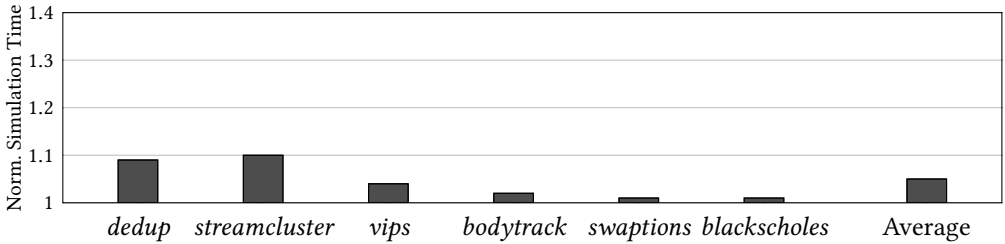
\begin{figure}[t]
\centering
    \centering
    \pgfplotstableread[col sep=comma]{data/overhead.csv}\datatable
    \pgfplotsset{
        overhead_axis/.style={
			ybar=0pt,
			bar width=15pt,
			ybar legend,
            width=12.5cm,
            height=2.5cm,
            xtick=data,
            enlarge x limits={true,abs value=0.6},
			major tick length=0,
            xticklabels from table={\datatable}{application},
            xticklabel style={font=\em\strut},
        },
        norm_sim_time/.style={
            ybar,
            other_bar,
            draw=black,
        },
    }
    \begin{tikzpicture}
        \begin{axis}[
            overhead_axis,
            ymajorgrids,
            ymin=1,
            ymax=1.4,
            xlabel={},
            ylabel={Norm. Simulation Time}]

            \plot[norm_sim_time] table[x=x,y=norm_sim_time] {\datatable};
        \end{axis}
    \end{tikzpicture}
    \caption{
        Simulation time normalized to HotSniper toolchain.
    }
    \label{fig:Parsec-Overhead}
\end{figure}

Compared to {\em HotSniper}, which runs core-only performance and thermal simulations, \comet executes thermal simulations for core and memory.  Figure~\ref{fig:Parsec-Overhead} compares simulation time for the PARSEC workloads running on a processor with off-chip 2D DRAM (\textit{2D-ext} core-memory configuration) under {\em HotSniper} and \comet. For \textit{2D-ext}, \comet runs separate thermal simulations for core and memory. Compared to {\em HotSniper}, we observe only a marginal increase in simulation time ($\sim$5\%, on average) using \comet. This observation is because the performance simulation is the dominant portion of the total simulation time and hence an additional thermal simulation leads to only a marginal increase. Furthermore, we simulated other configurations ({\em 3D-ext}, {\em 2.5D}, and {\em 3D-stacked}) and observed less than $\sim$2\% variation in simulation times. Overall, \comet leads to an acceptable increase of $\sim$5\% in simulation time to provide memory temperatures (in addition to core temperatures) at the epoch level.

\section{Conclusion and Future Work} 
\label{sec:conclusion}
High-performance high-density stacked core-memory configurations for multi-/many-core processors are becoming popular and need efficient thermal management. We present the first work featuring an integrated core and memory interval thermal simulation toolchain, namely \comet, supporting various core-memory configurations. \comet provides several useful features such as a thermal visualization (video), user-modifiable DTM policy, a built-in floorplan generator, easy simulation control, and an automated testing framework to facilitate system-level thermal management research for processors. We discussed various experimental studies performed using \comet, which will help researchers identify research opportunities and enable detailed, accurate evaluation of research ideas. Compared to a state-of-the-art core-only interval thermal simulation toolchain~\cite{pathania2018hotsniper}, \comet adds only an additional $\sim$5\% simulation-time overhead. The source code of \comet has been made open for public use under the {\em MIT} license.

\revision{
We plan to extend \comet to support 3D-stacked SRAM caches and NVM architectures.  We would also explore a plugin-based integration of thermal simulators to simplify the usage of emerging and more accurate thermal simulators with \comet. 
}
\ifdefined\IEEEformat
    \bibliographystyle{IEEEtran}
    \bibliography{ref}
\else
    \bibliographystyle{ACM-Reference-Format}
    \bibliography{ref}
\fi

%








\end{document}